\documentclass[prb,reprint,a4paper,superscriptaddress,showpacs,citeautoscript,floatfix]{revtex4-1}
\pdfoutput=1
\usepackage[charter]{mathdesign}
\usepackage{amsmath}
\usepackage[pdftex]{graphicx}
\usepackage{microtype}
\usepackage{bm}\let\vec\bm

\usepackage[svgnames]{xcolor}
\usepackage[
	colorlinks=True,linkcolor=DarkRed,citecolor=ForestGreen,urlcolor=MediumBlue,
	pdfstartview=FitH,bookmarks=False,pdfpagemode=UseNone
]{hyperref}

\def\strain{\vec{\varepsilon}}
\def\elDOS{N(E_{\mathrm{F}})}
\def\EEP{\langle\mathcal{J}^2\rangle}
\def\average{\langle\omega\rangle}
\renewcommand{\square}[1][]{\langle\omega^2_{#1}\rangle} 
\def\weight{w}

\def\anharmonicU{\tilde{U}}
\def\scaling{s}

\begin{document}

\title{\boldmath A theory of the strain-dependent critical field in Nb$_3$Sn,
based on anharmonic phonon generation}

\author{Davide Filippo Valentinis}
\affiliation{European Organization for Nuclear Research (CERN), 1211 Geneva 23,
Switzerland}
\affiliation{D{\'e}partement de Physique de la Mati{\`e}re Condens{\'e}e (DPMC),
University of Geneva, 24 quai Ernest-Ansermet, 1211 Geneva 4, Switzerland}
\author{Christophe Berthod}
\affiliation{D{\'e}partement de Physique de la Mati{\`e}re Condens{\'e}e (DPMC),
University of Geneva, 24 quai Ernest-Ansermet, 1211 Geneva 4, Switzerland}
\author{Bernardo Bordini}
\affiliation{European Organization for Nuclear Research (CERN), 1211 Geneva 23,
Switzerland}
\author{Lucio Rossi}
\affiliation{European Organization for Nuclear Research (CERN), 1211 Geneva 23,
Switzerland}

\date{January 13, 2014}

\begin{abstract}

We propose a theory to explain the strain dependence of the critical properties
in A15 superconductors. Starting from the strong-coupling formula for the
critical temperature, and assuming that the strain sensitivity stems mostly from
the electron-phonon $\alpha^2F$ function, we link the strain dependence of the
critical properties to a widening of $\alpha^2F$. This widening is attributed to
the nonlinear generation of phonons, which takes place in the anharmonic
deformation potential induced by the strain. Based on the theory of sum- and
difference-frequency wave generation in nonlinear media, we obtain an explicit
connection between the widening of $\alpha^2F$ and the anharmonic energy. The
resulting model is fit to experimental datasets for Nb$_3$Sn, and the anharmonic
energy extracted from the fits is compared with first-principles calculations.

\end{abstract}

\maketitle

\section{Introduction}
\label{sec:Introduction}

Nb$_3$Sn is one of the prominent materials for present and future applications
of superconductivity. It is used in high-field magnets for NMR spectroscopy
\cite{Miyazaki-1999}, for plasma confinement in nuclear fusion reactors
\cite{Mitchell-2009}, in focusing and beam-steering magnets for particle
accelerators, and is presently considered as a replacement for NbTi in the
future upgrade of the Large Hadron Collider magnets \cite{Bottura-2012}.
Nb$_3$Sn belongs to the crystalline symmetry group A15. It is a conventional
superconductor with a critical temperature $T_c$ around 17~K. The
superconducting transition is well explained by the phonon-mediated pairing
mechanism \cite{Arbman-1978, Klein-1979}, and the relatively high transition
temperature can be understood in terms of a strong electron-phonon coupling, and
a high density of states due to a narrow band of Nb $4d$ electrons at the Fermi
energy \cite{[{See, e.g., }] Paduani-2009}. The critical properties of Nb$_3$Sn
($T_c$, $B_{c2}$, $J_c$) vary considerably with the mechanical stress applied to
the material. As the electromagnetic forces scale with the square of the fields,
the strain dependence becomes increasingly important as the fields get higher.
With a strong sensitivity to strain, Nb$_3$Sn is an archetypical material to
investigate and model the strain dependence of the critical surface in
superconductors.

Previous investigations of the strain sensitivity of the critical parameters of
Nb$_3$Sn involved considerations associated with the structural transition from
cubic to tetragonal symmetry \cite{Gorkov-1976}. The progressive refinement of
empirical equations offered a description of uniaxial strain dependence
\cite{Ekin-1980, Taylor-2005}, as well as interpolation techniques for data
analysis \cite{tenHaken-1999, Godeke-2005}. Following pioneering hints by
Testardi \cite{Testardi-1971}, the role of anharmonicity as a source of the
strain dependence in Nb$_3$Sn was later recognized and emphasized by Marckievicz
\cite{Markiewicz-2004}.

Here, we propose a microscopic mechanism to explain the effect of strain-induced
anharmonicity on the phonon spectrum of Nb$_3$Sn. A new scaling law for the
strain dependence of the critical properties is deduced, based on the
Migdal-Eliashberg strong-coupling theory of superconductivity. This work has
been triggered by recent studies \cite{Bordini-2013}, showing that a relatively
simple exponential expression of critical field and current dependence on strain
[Eq.~(19) of Ref.~\onlinecite{Bordini-2013}] is capable of well fitting
experimental data of a large amount of samples, and over a large strain range.

The anharmonic terms in a crystalline potential induce interactions between
phonons. In a quasi-particle description, these processes correspond to
phonon-phonon scattering, and reduce the lifetime of the phonons. From a
wavelike perspective, they can be regarded as energy and momentum exchange
between coupled lattice vibrations. The latter conception, along with the
general theory of resonance in nonlinear systems, offers a complementary
insight: in a nonlinear medium, the mutual interaction of two coherent
propagating waves generates new waves with sum and difference frequencies, at
leading order in the nonlinearity. The new waves are amplified, and contribute
to the total spectrum of excitations. Here we argue that the same mechanism can
help understanding the effect of strain on phonons, and hence the detrimental
effect of strain on superconductivity. The mechanical stress induces---or
enhances---anharmonicity in the elastic potential. This favors the interaction
between lattice waves, generating new waves with sum and difference frequencies.
The phonon spectrum gets broadened, and the electron-phonon coupling is
accordingly reduced, leading to a decrease of the critical properties
\cite{Valentinis-2012}.

The simplest functional form which could describe the variation of the critical
properties under strain, based on the conjecture that these variations are
controlled by the strain-induced anharmonic energy $\anharmonicU$, is
$1/(1+A\anharmonicU)$, where $A$ is a constant. Since $\anharmonicU$ is an
increasing function of strain, this formula yields a bell-shaped curve as a
function of strain, consistently with experiments \cite{Bordini-2012,
Bordini-2013}. Our model yields the functional form
$1/\cosh[(B\anharmonicU)^{1/2}]^{\beta}$, with $B$ a constant, and $\beta$ a
number of order one. Both forms agree at small $\anharmonicU$, provided $A=\beta
B/2$, but the second form drops exponentially, rather than polynomially, at
large $\anharmonicU$. An exponential dependence in the strain turns out to give
a better description of the available experimental datasets, as emphasized in
Ref.~\onlinecite{Bordini-2013}.

The paper is organized as follows. The theory is presented in
Sec.~\ref{sec:theory}, and its application to Nb$_3$Sn in Sec.~\ref{sec:Nb3Sn}.
In Sec.~\ref{sec:strong-coupling}, we briefly review the strong-coupling theory
of superconductivity, emphasizing the role played by the width of the phonon
density of states. Section~\ref{sec:frequency-generation} deals with the effect
of anharmonicity on the phonon spectrum. The strain function and the resulting
model for the critical field are presented in Sec.~\ref{sec:strain-function}. In
Sec.~\ref{sec:Nb3Sn}, we first discuss the strain state in Nb$_3$Sn wires, and
propose a parametrization of the anharmonic energy
(Sec.~\ref{sec:pre-strained-wires}). We then present fits of the model to
experimental data, and compare the extracted anharmonic energy with
first-principles calculations (Sec.~\ref{sec:fits}). Our conclusions and
perspectives are summarized in Sec.~\ref{sec:conclusion}.
Appendices~\ref{app:elastic-constants} and \ref{app:sum-and-difference} contain
general considerations about nonlinear elasticity, our first-principles
calculations of the Nb$_3$Sn elastic constants, and a discussion of the
nonlinear generation of acoustic waves.

\section{Theory}
\label{sec:theory}

\subsection{Strong-coupling superconductivity and strain-dependent critical
temperature}
\label{sec:strong-coupling}

The strong-coupling theory of superconductivity extends the
Bardeen-Cooper-Schrieffer theory by including the dynamical structure of the
phonon-mediated pairing, in particular the retardation effects
\cite{Scalapino-1966}. The main ingredient of the theory is the electron-phonon
spectral function $\alpha^2F(\omega)$, which gives an average of the square of
the electron-phonon matrix element for electrons on the Fermi surface exchanging
phonons of frequency $\omega$. The numerical solution of the strong-coupling
equations for the critical temperature $T_c$ has been cast in a simple
analytical formula \cite{McMillan-1968, Dynes-1972, Allen-1975}, which depends
on a small number of physically meaningful parameters:
	\begin{equation}\label{eq:Tc}
		k_{\mathrm{B}}T_c=\frac{\hbar\average}{1.20}
		\exp\left[-\frac{1.04(1+\lambda)}{\lambda-\mu^*(1+0.62\lambda)}\right].
	\end{equation}
Here, $\average$ is of the order of the average phonon frequency, $\mu^*$ gives
the strength of the screened Coulomb repulsion on the Fermi surface
\cite{Morel-1962}, and $\lambda$ is a dimensionless parameter measuring the
strength of the electron-phonon interaction. The latter is related to the
$\alpha^2F$ function by
	\begin{equation}\label{eq:lambda-alphasquareF}
		\lambda=2\int_0^{\infty}d\omega\,\frac{\alpha^2F(\omega)}{\omega}.
	\end{equation}
This parameter also determines the renormalization of the Fermi velocity
$v_{\mathrm{F}}^*=v_{\mathrm{F}}/(1+\lambda)$ by the electron-phonon
interaction, as well as the enhancement of the electronic specific heat
coefficient $\gamma=(1+\lambda)\gamma_0$ with respect to the band value.
Alternatively, the coupling $\lambda$ may be written as \cite{McMillan-1968}
	\begin{equation}\label{eq:lambda-variance}
		\lambda=\frac{\elDOS\EEP}{M\square},
	\end{equation}
where $\elDOS$ is the electronic density of states (DOS) at the Fermi level,
$\EEP$ is the electronic contribution to the Fermi-surface average of the
squared electron-phonon matrix element, and $M$ is the average ionic mass. The
remaining parameter in Eq.~(\ref{eq:lambda-variance}), $\square$, will be our
main concern. It is defined as
	\begin{equation}\label{eq:variance}
		\square=\frac{2}{\lambda}\int_0^{\infty}d\omega\,\omega\,\alpha^2F(\omega).
	\end{equation}
For a single Einstein phonon of frequency $\omega_0$, we have
$\square=\omega_0^2$. For a Debye spectrum, assuming that $\alpha^2F$ is simply
proportional to the quadratic phonon DOS, we find
$\square=\omega_{\mathrm{D}}^2/2$, where $\omega_{\mathrm{D}}$ is the Debye
frequency. Thus, for a general phonon spectrum, $\square^{1/2}$ gives a measure
of the width of the phonon DOS.

The central assumption of the present study is that the strain-induced variation
of the critical properties in Nb$_3$Sn and similar compounds is dominated by
changes in the phonon spectrum. We accordingly infer that the purely electronic
quantities, namely $\mu^*$, $\elDOS$, and $\EEP$ in Eqs.~(\ref{eq:Tc}) and
(\ref{eq:lambda-variance}), can be considered strain-independent in a first
approximation. Recent first-principles calculations for Nb$_3$Sn suggest that
the electron and phonon spectra both vary with the applied strain
\cite{DeMarzi-2012}. So far, however, the question whether the strain
sensitivity in A15 superconductors is due mainly to lattice or to electronic
degrees of freedom has not been settled, neither experimentally nor
theoretically \cite{Taylor-2005, Oh-2006, Godeke-2006}. The justification for
the present approach comes primarily from its ability to describe experimental
data.

\begin{table}[b]
\caption{\label{tab:lambda-mu}
Electron-phonon coupling $\lambda$, and screened Coulomb repulsion $\mu^*$, for
some A15 superconductors.
}
\begin{tabular*}{\columnwidth}{l@{\extracolsep{\fill}}ccl}
\hline\hline
Material & $\lambda$ & $\mu^*$ & Ref.\\
 \hline
Nb$_3$Sn & 1.80 & 0.16 & \onlinecite{Wolf-1980}\\
Nb$_3$Al & 1.70 & 0.15 & \onlinecite{Kwo-1981}\\
Nb$_3$Ge & 1.64 & 0.12 & \onlinecite{Kihlstrom-1984}
\footnote{Values for the sample with the highest $T_c$}\\
V$_3$Si  & 1.07 & 0.13 & \onlinecite{Delaire-2008}\\
\hline\hline
\end{tabular*}
\end{table}

We propose that the interactions between phonons, generated by the anharmonic
terms in the elastic energy, produce secondary vibrational modes with sum and
difference frequencies, as discussed in more detail in the next section. This
process leads to a broadening of the phonon spectrum, hence to an increase of
$\square$. A change in $\average$ may also result, but this change is expected
to be small compared with the change of $\square$, due to cancellations between
sum and difference frequencies. Moreover, since $\square$ goes into the
exponential in the formula (\ref{eq:Tc}), while $\average$ enters only as a
pre-factor, we will neglect the variation of $\average$ hereafter for
simplicity. Another simplification arises, because the value of the screened
Coulomb repulsion is generally small with respect to $\lambda$ in A15 compounds.
A few representative values are listed in Table~\ref{tab:lambda-mu}. Neglecting
terms of order $\mu^*/\lambda$ in the square brackets in Eq.~(\ref{eq:Tc}), we
form the ratio between the critical temperature $T_c(\strain)$ in the presence
of a strain described by the tensor $\strain$, and the value $T_c(0)$ at
equilibrium:
	\begin{equation}\label{eq:Tc-lambda}
		\frac{T_c(\strain)}{T_c(0)}=\exp\left[-\frac{1.04}{\lambda(0)}
		\left(\frac{\lambda(0)}{\lambda(\strain)}-1\right)\right].
	\end{equation}
Considering Eq.~(\ref{eq:lambda-variance}), we furthermore see that
$\lambda(0)/\lambda(\strain)=\square[\strain]/\square[0]$, where $\square[0]$
(respectively, $\square[\strain]$) is the value of the parameter $\square$ at
equilibrium (respectively, under strain). We conclude that, according to the
hypotheses made so far, the strain-induced variation of $T_c$ depends on the
relative variation of the phonon-spectrum ``width'', more precisely the
parameter $\square$.

\subsection{Broadening of the phonon spectrum by nonlinear wave generation}
\label{sec:frequency-generation}

As stated in the Introduction, the interaction between phonons can be viewed, in
a quasi-particle picture, as phonon-phonon scattering processes, which reduce
the lifetime and the mean-free path of the phonons. Alternatively, it may be
regarded as the generation of secondary waves in a nonlinear medium. In both
views, the interaction has the effect of widening the phonon spectrum of a
hypothetical perfect harmonic crystal, either by broadening individual phonon
lines in the former view, or by adding additional phonon modes in the
latter---which is the preferred view in the present study. Phonon-phonon
interactions are present in materials even at equilibrium, due to intrinsic
anharmonic terms in the elastic deformation potential. The application of a
stress, with the associated deformation of the unit cell, reinforces anharmonic
contributions by driving the crystal away from its equilibrium point. The width
of the electron-phonon spectral function $\alpha^2F(\omega)$, and consequently
the parameter $\square$, are expected to increase with the applied strain,
reflecting the increased interaction among phonons. We may introduce two
limiting values, $\square[0]$ and $\square[1]$, corresponding respectively to
the equilibrium state, and to a high-strain state where the generation of
secondary phonons with sum and difference frequencies would have exhausted all
pairs of primary phonons. Between these two limits, we parametrize the variation
of $\square$ with strain as
	\begin{equation}\label{eq:square-strain}
		\square[\strain]-\square[0]=\weight(\strain)\left(\square[1]-\square[0]\right),
	\end{equation}
where, by assumption, the anharmonic weight factor $\weight(\strain)$ is
proportional to the total amplitude of the secondary waves generated by the
strain-induced anharmonic terms in the elastic energy.

In order to build a model for $\weight(\strain)$, we borrow idea familiar in
nonlinear optics \cite{Boyd-2003}. Two electromagnetic waves with frequencies
$\omega_1$ and $\omega_2$, entering a nonlinear resonant cavity, generate
secondary waves with frequencies $\omega_3=\omega_1\pm\omega_2$, at first order
in the nonlinearity. The wave generation obeys energy conservation rules, and
phase matching conditions for the momenta \cite{Boyd-2003}. The amplitudes of
the primary waves progressively die out, while their energy is transferred to
the secondary waves, whose amplitudes grow. A very similar phenomenon occurs for
acoustic waves in a nonlinear elastic medium. The case of longitudinal waves is
presented in Appendix~\ref{app:sum-and-difference}. We find that the amplitude
$A_3$ of the secondary waves increase in time according to
$\omega_3\,A_3(t)=(\tilde{U}/\rho)^{1/2}\tanh[tq_{12}(\tilde{U}/\rho)^{1/2}]$,
where $\tilde{U}$ is the energy transferred from the primary to the secondary
waves, $\rho$ is the mass density of the medium, and
$q_{12}=|b/2|\sqrt{q_1q_2}$, with $b$ a nonlinear factor [Eq.~(\ref{eq:b})], and
$q_{1,2}$ the wave numbers of the primary waves. If the primary waves interact
coherently with a negligible damping during a time $\tau$, the total amplitude
generated during the coherence time is therefore
	\begin{equation}
		\int_0^{\tau}dt\,A_3(t)\propto\ln\left[\cosh
		\left(\sqrt{\tilde{U}/\rho}q_{12}\tau\right)\right],
	\end{equation}
where the pre-factor is independent of the energy $\tilde{U}$. As the weight
factor $\weight(\strain)$ should be dimensionless, and proportional to the
amplitude of generated waves, it is natural to take
	\begin{equation}\label{eq:weight-strain}
		\weight(\strain)=\ln\left[\cosh\left(
		\sqrt{\anharmonicU(\strain)/\rho}\,Q\tau\right)\right].
	\end{equation}
$\anharmonicU(\strain)$ is the strain-induced anharmonic energy available for
the generation of secondary phonons, and $Q=|b/2|\langle\sqrt{q_1q_2}\rangle$ is
a typical phonon momentum. Considering that the primary momenta $q_{1,2}$ vary
between 0 and $\omega_{\mathrm{D}}/v$, with $v$ the velocity of sound, we find
in three dimensions
$\langle\sqrt{q_1q_2}\rangle=(36/49)\omega_{\mathrm{D}}/v\approx(3/4)
\omega_{\mathrm{D}}/v$. Finally, $\tau$ plays the role of a phonon coherence
time. At low temperature, it is given by the average phonon lifetime, while at
high temperature it is cut by the thermal limit $h/(k_{\mathrm{B}}T)$. Equation
(\ref{eq:weight-strain}) correctly gives $\weight(0)=0$, as well as
$w(\strain)\equiv0$ in the absence of anharmonicity ($b=0$).

\subsection{Exponential strain function and critical field}
\label{sec:strain-function}

The set of equations (\ref{eq:weight-strain}), (\ref{eq:square-strain}),
(\ref{eq:Tc-lambda}), and (\ref{eq:lambda-variance}) lead, after a simple
algebra, to the expression
	\begin{equation}\label{eq:scaling-Tc}
		\frac{T_c(\strain)}{T_c(0)}=\cosh\left(\sqrt{\anharmonicU(\strain)/\rho}\,Q\tau
		\right)^{-\frac{1.04}{\lambda(0)}\left(\frac{\square[1]}{\square[0]}-1\right)}.
	\end{equation}
As the exponent is negative, and $\anharmonicU(\strain)$ is expected to increase
monotonically with increasing strain, $T_c(\strain)$ is expected to be a
monotonically decreasing function of strain according to
Eq.~(\ref{eq:scaling-Tc}). It is customary to introduce a ``strain function''
$\scaling(\strain)$ in order to describes the strain dependence of the upper
critical field at zero temperature:
	\begin{equation}\label{eq:scaling-Bc2}
		\scaling(\strain)=\frac{B_{c2}(\strain)}{B_{c2}(0)}.
	\end{equation}
Experimentally, it was found \cite{Ekin-1980} that the strain dependence of
$T_c$ scales with that of $B_{c2}$, i.e.,
$T_c(\strain)/T_c(0)=[B_{c2}(\strain)/B_{c2}(0)]^{1/\alpha}$ with $\alpha\approx
3$. This leads to the following model for the strain function:
	\begin{equation}\label{eq:strain-function}
		\scaling(\strain)=\cosh\left(\sqrt{\anharmonicU(\strain)/\rho}\,Q\tau
		\right)^{-\frac{1.04\alpha}{\lambda(0)}\left(\frac{\square[1]}{\square[0]}-1\right)}.
	\end{equation}
Equation~(\ref{eq:strain-function}) is our main result. It shows that the strain
dependence of $\scaling(\strain)$ stems from the strain-induced anharmonic
contributions to the elastic energy. The temperature dependence of the critical
field is described by the scaling law \cite{deGennes-1999}
$B_{c2}(T)/B_{c2}(0)=(1-T/T_c)^{\gamma}$, with an exponent $\gamma\approx1.5$.
We can therefore write the temperature- and strain-dependent critical field in
the form
	\begin{equation}\label{eq:critical-field}
		B_{c2}(T,\strain)=B_{c20}\left\{1-\left(\frac{T}{T_{c0}}\right)^{\gamma}
		[\scaling(\strain)]^{-\frac{\gamma}{\alpha}}\right\}
		\scaling(\strain),
	\end{equation}
where $B_{c20}$ is the zero-temperature and zero-strain upper critical field,
and $T_{c0}$ is the zero-strain critical temperature. In the next section, we
use Eqs.~(\ref{eq:strain-function}) and (\ref{eq:critical-field}) to fit
critical-field data measured on Nb$_3$Sn wires, extract the anharmonic energy
$\anharmonicU(\strain)$, and compare with the anharmonic energy calculated from
first principles for bulk Nb$_3$Sn strained like in the wires.

\section{\boldmath Application to Nb$_3$Sn superconducting wires}
\label{sec:Nb3Sn}

\subsection{Anharmonic energy of pre-strained Nb$_3$Sn}
\label{sec:pre-strained-wires}

In order to compare the model (\ref{eq:strain-function}) and
(\ref{eq:critical-field}) with experimental determinations of
$B_{c2}(T,\strain)$, we have to specify the strain, and to parametrize the
behavior of $\anharmonicU(\strain)$. Following Ref.~\onlinecite{tenHaken-1994},
we introduce three invariants $I_1$, $J_2$, and $J_3$ to represent the
deformation. $I_1$ is the hydrostatic invariant, given by
$I_1=\varepsilon_1+\varepsilon_2+\varepsilon_3$, where $\varepsilon_i$ are the
values of the strain along the three principal directions. $J_2$ and $J_3$ are
the second and third deviatoric invariants, defined as
$J_2=[(\varepsilon_1-\varepsilon_2)^2+(\varepsilon_2-\varepsilon_3)^2+
(\varepsilon_3-\varepsilon_1)^2]/6$, and
$J_3=(\varepsilon_1-I_1/3)(\varepsilon_2-I_1/3)(\varepsilon_3-I_1/3)$. In a
harmonic crystal, the elastic energy is quadratic, and takes a simple form in
terms of the two second-order invariants $I_1^2$ and $J_2$,
	\begin{equation}\label{eq:harmonic_energy}
		U_{\mathrm{harm}}=\frac{1}{6}\frac{E}{1-2\nu}I_1^2+\frac{E}{1+\nu}J_2,
	\end{equation}
where $E$ is the Young modulus and $\nu$ is the Poisson ratio (see
Appendix~\ref{app:elastic-constants}). This harmonic energy does not contribute
to the interaction between phonons. Under the effect of stress, additional terms
of higher orders in the strain contribute to the elastic energy. Terms of third
order can be proportional to $I_1^3$, $I_1J_2$, and $J_3$, while terms of fourth
order include the combinations $I_1^4$, $I_1^2J_2$, $I_1J_3$, and $J_2^2$. Our
goal is to represent the anharmonic energy---an increasing function of
strain---with as few parameters as possible. We therefore exclude contributions
which are not positive-definite as a function of strain. With the strain
configuration of interest to us, namely $\varepsilon_2=\varepsilon_3$ (see
below), this excludes all third-order terms, as well as the fourth-order term
$I_1J_3$: all of them are odd functions of $\varepsilon_1+2\varepsilon_2$ and/or
$\varepsilon_1-\varepsilon_2$. Our parametrization of the anharmonic energy is
therefore
	\begin{equation}\label{eq:Utilde}
		\anharmonicU(\strain)=p_1^{ }\,I_1^4+p_{12}^{ }\,I_1^2J_2^{ }+p_2^{ }\,J_2^2.
	\end{equation}
The three parameters are nonnegative, and have the unit of energy per unit
volume, i.e., pressure. One further constraint will emerge from the analysis of
our particular strain model.

Here we focus on the strain experienced by Nb$_3$Sn strands during
critical-current measurements under applied longitudinal strain. In order to
describe these strain conditions, we use the same assumptions as in
Ref.~\onlinecite{Bordini-2013}: the principal strain component $\varepsilon_1$
is along the strand axis; $\varepsilon_1=\varepsilon_{l0}+\varepsilon_a$;
$\varepsilon_2=\varepsilon_3=\varepsilon_{t0}-\nu\varepsilon_a$, where
$\varepsilon_{l0}$ and $\varepsilon_{t0}$ are the initial longitudinal and
transverse strains, respectively, and $\nu$ is the effective poisson ratio,
for which we use the value $\nu=0.36$ as in Ref.~\onlinecite{Bordini-2013}.
Rewriting the pre-strain as the sum of an
hydrostatic part $\varepsilon_0$, and a deviatoric part $\varepsilon_{10}$,
these relations become
$\varepsilon_1=\varepsilon_0+\varepsilon_{10}+\varepsilon_a$, and
$\varepsilon_2=\varepsilon_3=\varepsilon_0-\nu\varepsilon_{10}-\nu\varepsilon_a$.

Upon varying $\varepsilon_a$, the critical field goes through a maximum at a
strain close to the value $\varepsilon_a^*$, which restores the cubic symmetry
of the superconductor \cite{tenHaken-1999, Muzzi-2012}. In the strain
configuration described above, we have $\varepsilon_a^*=-\varepsilon_{10}$,
which implies $I_1(\varepsilon_a^*)=3\varepsilon_0$, and
$J_2(\varepsilon_a^*)=0$. Therefore, the anharmonic energy
$\anharmonicU(\varepsilon_a^*)=p_1(3\varepsilon_0)^4$ accounts for the reduction
of the critical field under purely hydrostatic strain. The consistency of the
model requires that the function $\anharmonicU(\strain)$ as a function of
$\varepsilon_a$ has a single minimum close to $\varepsilon_a=\varepsilon_a^*$,
and otherwise varies monotonically. By constraining the parameters to satisfy
$p_2/p_{12}>\theta(1-4\theta p_1/p_{12})/(1+12\theta p_1/p_{12})$, with
$\theta=(3/2)[(1-2\nu)/(1+\nu)]^2$, we ensure that the second derivative of
$\anharmonicU(\strain)$ with respect to $\varepsilon_a$ is positive for any
strain.

Before turning to the fits, we note that the parametrization (\ref{eq:Utilde})
may not be the best choice in every situation. In the case of pre-strained
wires, the formula (\ref{eq:Utilde}) generates all powers of $\varepsilon_a$
from zero to four, and has enough freedom to grant a good fit. For hydrostatic
pressure, however, Eq.~(\ref{eq:Utilde}) only contains the fourth power of the
strain. In this case, it may prove necessary to retain a term proportional to
$I_1^3$.

\subsection{Fits to critical-field data, and discussion of the anharmonic energy}
\label{sec:fits}

With the parametrization (\ref{eq:Utilde}), and enforcing the inequality on
$p_2/p_{12}$, the strain function (\ref{eq:strain-function}) can be recast in
the form
	\begin{subequations}\label{eq:scaling-model}
	\begin{align}
		\label{eq:scaling-model1}
		\scaling(\strain)&=\left[\cosh\left(\!
		C_0\sqrt{C_1I_1^4+I_1^2J_2^{ }+(\delta+C_2)\,J_2^2}\right)\right]^{-\beta}\\
		I_1&=(1-2\nu)(\varepsilon_a-\varepsilon_a^*)+3\varepsilon_0\\
		J_2&=(1/3)(1+\nu)^2(\varepsilon_a-\varepsilon_a^*)^2\\
		\delta&=\theta(1-4\theta C_1)/(1+12\theta C_1)\\
		\theta&=(3/2)[(1-2\nu)/(1+\nu)]^2.
	\end{align}
	\end{subequations}
The positive dimensionless parameters $C_0$, $C_1$, and $C_2$ are related to the
parameters giving the anharmonic energy (\ref{eq:Utilde}) by
$p_{12}=\rho[C_0/(Q\tau)]^2$, $p_1=C_1p_{12}$, and $p_2=(\delta+C_2)p_{12}$. The
exponent $\beta$ depends on the electron-phonon coupling $\lambda(0)$, and on
the broadening of the phonon spectrum $\square[1]/\square[0]$, as can be seen
from Eq.~(\ref{eq:strain-function}). An estimate of this exponent for Nb$_3$Sn
may be obtained as follows. The $\alpha^2F$ function determined by inverting
tunneling data in Ref.~\onlinecite{Freericks-2002} has four main features, at
$4.5$, $8.0$, $18.6$, and $24.9$~meV. A very crude way of computing $\square[0]$
is to use
$\square[0]\approx\big(\sum_i\omega_i\big)/\big(\sum_i\omega_i^{-1}\big)$, where
the sums run over the four characteristic frequencies. The result is
$\square[0]^{1/2}\approx 11.3$~meV. In order to check the reliability of this
approximation, we take the complete $\alpha^2F$ function from
Ref.~\onlinecite{Freericks-2002}, and perform the integrals in
Eqs.~(\ref{eq:lambda-alphasquareF}) and (\ref{eq:variance}). This gives the
values $\lambda=2.74$ and $\square^{1/2}=12.6$~meV. The latter exact value is
quite close to the approximate result 11.3~meV. This suggest to compute
$\square[1]$ by the same method: from the set of characteristic frequencies, we
build the set of sum and difference frequencies, which together give
$\square[1]^{1/2}\approx15.8$~meV. Using $\alpha=3$ and $\lambda(0)=2.74$, we
finally deduce $\beta=1.09$.

The behavior of $s(\strain)$ close to the point $\varepsilon_a^*$ may be studied
by expanding Eq.~(\ref{eq:scaling-model}) as
$\scaling(\strain)=1\big/\big[1+\sum_ia_i(\varepsilon_a-\varepsilon_a^*)^i\big]$
. The zeroth order coefficient is $a_0=(81/2)\beta C_0^2C_1^{ }\varepsilon_0^4$,
to leading order in $\varepsilon_0$. This confirms that, according to the model
(\ref{eq:scaling-model}), the critical field decreases as $\varepsilon_0^4$
under hydrostatic strain. Up to fourth order, the coefficients $a_i$ involve the
combinations $\beta C_0^2$, $\beta C_0^2C_1^{ }$, and $\beta C_0^2C_2^{ }$.
Therefore, only three of the four parameters $\beta, C_{0,1,2}$ can be
determined by fitting experimental data close to
$\varepsilon_a=\varepsilon_a^*$: variations of the parameter $\beta$ can be
absorbed into a redefinition of $C_0$. In view of this, we shall adopt the value
$\beta=1$ in the following, when fitting (\ref{eq:critical-field}) and
(\ref{eq:scaling-model}) to experimental data.

Fitting the model to the Furukawa dataset of Ref.~\onlinecite{Bordini-2013}, we
find that $C_2=0$, which implies that the inequality imposed on $p_2/p_{12}$ is
saturated. A better fit could thus be achieved by allowing $C_2<0$, but the
resulting strain function would be non-monotonic, and show an unphysical
increase at large compressive strain. We fixed $C_2$ to zero for fitting all
datasets. We also find that $C_1\sim 10^{-2}$, and that $C_1$ and $B_{c20}$ are
correlated. This can be understood, since the value of the critical field at
$\varepsilon_a^*$ is
$B_{c2}(T,\varepsilon_a^*)=B_{c20}[1-(81/2)C_0^2C_1\varepsilon_0^4]$, up to
terms of order $(T/T_{c0})^{\gamma}$ and $\varepsilon_0^8$. The value of
$B_{c2}(T,\varepsilon_a^*)$ can be reproduced by fixing one of the parameters
$B_{c20}$ or $C_1$, and adjusting the other, without affecting significantly the
$B_{c2}(T,\varepsilon_a)$ curve. We have chosen to fix $C_1=0.035$, such that
the fitted value of $B_{c20}$ is 28.6~T, as in Ref.~\onlinecite{Bordini-2013}. A
theoretical estimate of $C_1$ can be obtained from first-principles
calculations, as described in Appendix~\ref{app:elastic-constants}. The
Murnaghan formula for the elastic energy is of third order in the quadratic
strain invariants. It can be used to derive a fourth-order expression in the
linear invariants $I_1$, $J_2$, and $J_3$ [Eq.~(\ref{eq:elastic-energy-2})]. In
the latter expression, the ratio between the coefficients of the $I_1^4$ and
$I_1^2J_2^{ }$ terms corresponds to $C_1$. Using the equilibrium theoretical
Lam{\'e} and Murnaghan coefficients given in
Appendix~\ref{app:elastic-constants}, we thus obtain $C_1=0.048$. With the
effective coefficients which fit the elastic energy in the strain configuration
of the Furukawa dataset (see below, the discussion of Fig.~\ref{fig:fig2}), we
find $C_1=0.032$, in excellent agreement with the value $0.035$ used in the
fits.

With $C_1$ and $C_2$ fixed to $0.035$ and $0$, respectively, the four adjustable
parameters of the model are $B_{c20}$, $\varepsilon_0$, $\varepsilon_a^*$, and
$C_0$. $\varepsilon_a^*$ is close to the strain at which the critical field
reaches a maximum; it is therefore strongly constrained by the data, and has an
immediate physical interpretation as $-\varepsilon_{10}$, i.e., minus the
longitudinal deviatoric pre-strain. $\varepsilon_0$ controls the asymmetry of
the strain function with respect to $\varepsilon_a=\varepsilon_a^*$: the
function is symmetric if $\varepsilon_0=0$, a property also verified by the
exponential strain function of Ref.~\onlinecite{Bordini-2013}. This shows that
the asymmetry is due to the hydrostatic pre-compression of the superconductor.
The measurements of Ref.~\onlinecite{Muzzi-2012} support this claim: more
symmetric scaling curves are observed for bare wires than for jacketed wires.
The asymmetry of the experimental strain function is a characteristic change to
a weaker slope, appearing under tensile strain in wires which have a compressive
hydrostatic pre-strain. In all datasets of Ref.~\onlinecite{Bordini-2013}
showing a change of slope, the latter is seen under compressive strain,
suggesting that the superconductor in the strands was subjected to a tensile
hydrostatic pre-strain. The strain function (\ref{eq:scaling-model}) displays a
change of slope at $\varepsilon_a\approx\varepsilon_a^*-\varepsilon_0/(1-2\nu)$.
Consistently with the experiments, the change of slope occurs on the tensile
side if $\varepsilon_0<0$, and on the compressive side if $\varepsilon_0>0$,
because $\nu<1/2$.

\begin{figure}[tb]
\includegraphics[width=0.9\columnwidth]{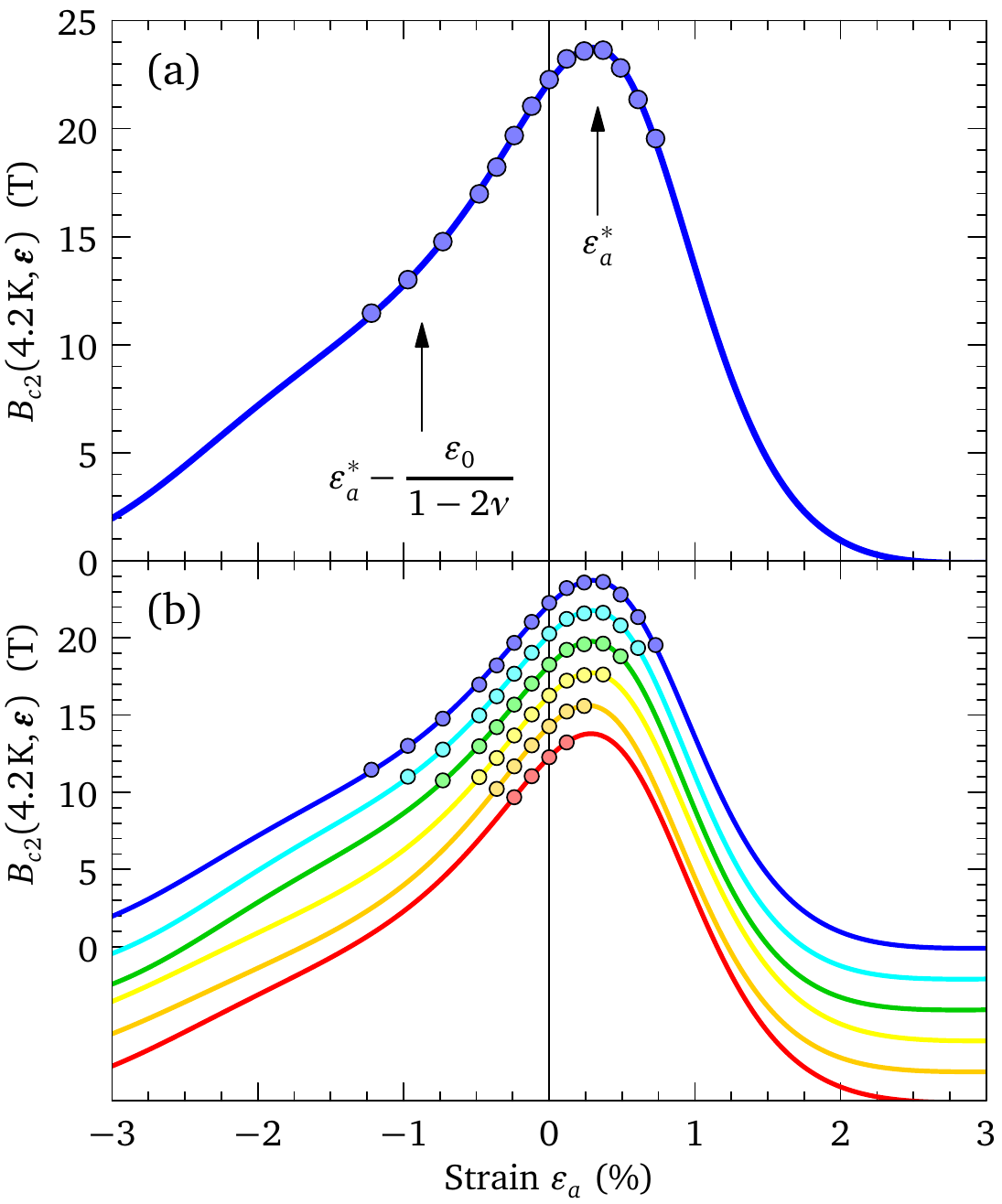}
\caption{\label{fig:fig1}
Fit of the model (\ref{eq:critical-field}) and (\ref{eq:scaling-model}), to the
Furukawa dataset of Ref.~\onlinecite{Bordini-2013}. The following model
parameters were fixed (see text): $\alpha=3$, $\gamma=1.5$, $T=4.2$~K,
$T_{c0}=17$~K, $\nu=0.36$, $\beta=1$, $C_1=0.035$, $C_2=0$. In panel (a), the
whole dataset was used, and the resulting fitted parameters are
$B_{c20}=28.6$~T, $\varepsilon_0=0.34\%$, $\varepsilon_a^*=0.33\%$, and
$C_0=1.70$. In panel (b), subsets of the data were considered: each curve was
fitted to the subset of points shown on top of it. The curves are offset
vertically by 2~T for clarity.
}
\end{figure}

Figure~\ref{fig:fig1}(a) shows a fit of the model to the Furukawa dataset of
Ref.~\onlinecite{Bordini-2013}. The resulting positive value
$\varepsilon_0=0.34\%$ can be understood, since the change of slope occurs on
the compressive side. As expected, the cubic point $\varepsilon_a^*$ is close to
where the maximum occurs. The initial longitudinal strain
$\varepsilon_{l0}=\varepsilon_0-\varepsilon_a^*$ is almost zero, but the
transverse pre-strain $\varepsilon_{t0}=\varepsilon_0+\nu\varepsilon_a^*=0.46\%$
is tensile. The fitted curve is robust if fewer data points are used, as
demonstrated in Fig.~\ref{fig:fig1}(b). The values obtained for $\varepsilon_0$
and $\varepsilon_a^*$, in particular, vary by no more than $\sim10\%$ when data
points are removed.

Similar fits can be achieved to the other datasets reported in
Ref.~\onlinecite{Bordini-2013}. The resulting parameters are collected in
Table~\ref{tab:fits}. In all cases, we find a positive value of
$\varepsilon_0\sim0.3$--$0.4\%$, indicating that the superconductor in the
strands was subject to a tensile hydrostatic pre-strain. The net longitudinal
pre-strain $\varepsilon_{l0}$ is close to zero for the wires prepared by
following the bronze route, and positive for the others. We note that the weight
factor $w(\strain)$ remains smaller than unity in all fits, as needed for the
consistency of Eq.~(\ref{eq:square-strain}).

\begin{table}[tb]
\caption{\label{tab:fits}
Parameters resulting from fitting the model (\ref{eq:critical-field}) and
(\ref{eq:scaling-model}) to the datasets of Ref.~\onlinecite{Bordini-2013},
including strands fabricated using the bronze route (BR), the internal-tin
method (IT), and the powder-in-tube method (PIT). The fixed model parameters are
as in Fig.~\ref{fig:fig1}. The last column gives the root-mean square deviation.
Note that the constants $C_i$ refer to strain values expressed in percentages.
}
\begin{tabular*}{\columnwidth}{l@{\extracolsep{\fill}}cccccc}
\hline\hline
Strand & Type & $B_{c20}$ (T) & $\varepsilon_0$ (\%) & $\varepsilon_a^*$ (\%) & $C_0$ & RMS (T) \\[1mm]
\hline
Furukawa & BR  & 28.6 & 0.34 & 0.33 & 1.70 & 0.10 \\
VAC      & BR  & 28.4 & 0.30 & 0.35 & 1.99 & 0.06 \\
OKSC     & IT  & 28.2 & 0.29 & 0.11 & 2.12 & 0.02 \\
OST      & IT  & 28.8 & 0.40 & 0.14 & 1.40 & 0.07 \\
PORI     & IT  & 28.6 & 0.39 & 0.13 & 1.45 & 0.12 \\
U.G.8305 & BR  & 28.6 & 0.33 & 0.32 & 1.40 & 0.08 \\
U.G.7567 & IT  & 29.3 & 0.40 & 0.29 & 1.33 & 0.04 \\
U.G.0904 & PIT & 31.1 & 0.34 & 0.21 & 1.63 & 0.05 \\
\hline\hline
\end{tabular*}
\end{table}

The data reported in Refs.~\onlinecite{Mondonico-2010, Muzzi-2012} for the
jacketed wires are the only ones, to our knowledge, which present a change of
slope for tensile strain. This change of slope occurs at
$\varepsilon_a\sim0.83\%$ in a wire where the largest critical current is
observed at $\varepsilon_a\sim0.55\%$. In these experiments, the lattice
parameters were measured as a function of strain. In this particular wire, the
maximum of the critical current coincides with the point $\varepsilon_a^*$ of
cubic symmetry (see Fig.~11 of Ref.~\onlinecite{Muzzi-2012}). The measured
lattice parameter at the cubic point is $5.276$~\AA, while the reported
equilibrium parameter is $5.28$~\AA. Thus the hydrostatic pre-strain is
$\varepsilon_0\approx-0.076\%$. According to our model, the change of slope is
expected at a strain $-\varepsilon_0/(1-2\nu)$ measured from the position of the
maximum, as indicated in Fig.~\ref{fig:fig1}(a), that is, at
$\varepsilon_a\sim0.82\%$, in very good agreement with the observations.

We now turn to the discussion of the anharmonic energy. The core idea of the
model is that the strain dependence of the critical properties is governed by
the anharmonic contributions to the elastic energy
[Eq.~(\ref{eq:strain-function})]. Since the parametrization (\ref{eq:Utilde})
provides a good fit, it allows one to extract the function $\tilde{U}(\strain)$
from experimental data, in view of a comparison with an independent
determination of the anharmonic energy. We have calculated the elastic energy of
bulk Nb$_3$Sn from first-principles, in a strain configuration as determined for
the Furukawa dataset in Table~\ref{tab:fits} (see
Appendix~\ref{app:elastic-constants} for details). The result is shown as a
function of applied strain in Fig.~\ref{fig:fig2}. In order to extract the
anharmonic terms, we fitted these data using Eq.~(\ref{eq:elastic-energy-1}).
Such a fit can not determine reliably the five elastic constants. We therefore
fixed the Young modulus and Poisson ratio to the values $E=130$~GPa and
$\nu=0.36$ measured on Nb$_3$Sn strands, and fitted the three Murnaghan
coefficients $\ell$, $m$, and $n$. The resulting fit is the thin solid line in
Fig.~\ref{fig:fig2}, and yields a nonlinear factor $b=23$ [Eq.~(\ref{eq:b})],
and an estimate $C_1=0.032$, as indicated previously. Setting the Murnaghan
coefficients to zero, we obtain the quadratic energy shown as a thin dashed
line. (Notice that this function is not exactly quadratic in $\varepsilon_a$,
due to the pre-strain.) The difference is the anharmonic energy, shown as a
thick dashed line.

\begin{figure}[tb]
\includegraphics[width=0.7\columnwidth]{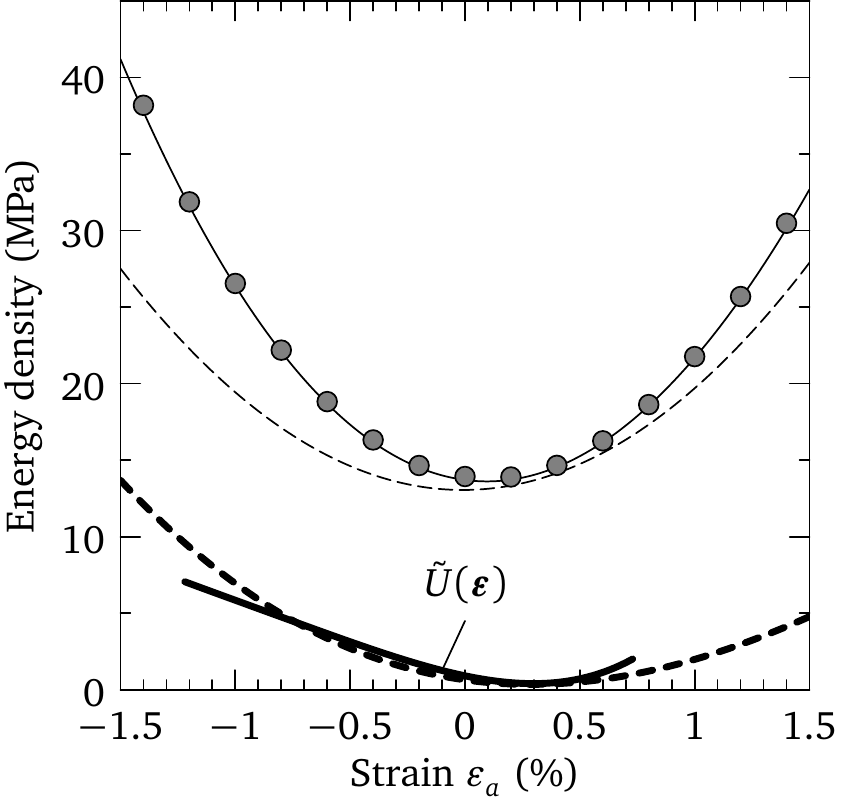}
\caption{\label{fig:fig2}
Elastic and anharmonic energies of strained Nb$_3$Sn. The dots are the result of
first-principles calculations for the strain configuration of the Furukawa
dataset in Table~\ref{tab:fits}. The thin solid line is a fit to
Eq.~(\ref{eq:elastic-energy-1}), which contains harmonic (thin dashed) and
anharmonic (thick dashed) contributions. The thick solid line is the anharmonic
energy (\ref{eq:Utilde1}) for $b=23$ and $\tau=0.9$~ps, in the range of strain
where experimental data is available. 
}
\end{figure}

Using the expression of $Q$ given near the end of
Sec.~\ref{sec:frequency-generation}, the relation between $\rho v^2$ and the
Young modulus and Poisson ratio (Appendix~\ref{app:elastic-constants}), and
setting $C_2=0$, the anharmonic energy (\ref{eq:Utilde}) becomes
	\begin{equation}\label{eq:Utilde1}
		\tilde{U}(\strain)=\frac{E}{1-2\nu}\frac{1-\nu}{1+\nu}\left(\frac{8/3}
		{b\omega_{\mathrm{D}}\tau}\right)^2C_0^2
		\left(C_1^{ }\,I_1^4+I_1^2J_2^{ }+\delta\,J_2^2\right).
	\end{equation}
We take $E=130$~GPa and $\nu=0.36$ as above, the parameters of the Furukawa
dataset, and $\omega_{\mathrm{D}}=31$~THz, corresponding to a Debye temperature
of 234~K (Ref.~\onlinecite{Guritanu-2004}). With $b$ fixed to the value 23
quoted above, we adjust $\tau$ such that the order of magnitude of
$\tilde{U}(\strain)$ agrees with the anharmonic energy calculated from first
principles: this leads to $\tau\sim0.9$~ps (Fig.~\ref{fig:fig2}). Repeating the
analysis for all datasets in Table~\ref{tab:fits}, we consistently find values
of $b$ of the order of 20, and values of $\tau$ between 0.8 and 1.3~ps. These
values are smaller than the thermal limit $h/(k_{\mathrm{B}}T)=11$~ps at
$4.2$~K. However, they are comparable to the lifetime of acoustic phonons in
Nb$_3$Sn. In the superconducting state, the width of the acoustic phonons lines
measured by inelastic neutron scattering \cite{Axe-1973} is typically
$2\Gamma=0.6$--$1.2$~meV. The corresponding phonon lifetime is
$\tau_{\text{ph}}=\hbar/(2\Gamma)=0.55$--$1.1$~ps. We conclude that the phonon
lifetime is the limiting factor in the nonlinear generation of sum and
difference-frequency phonons. The nice agreement between $\tau$ as determined
from the fits and $\tau_{\text{ph}}$ is an encouraging consistency check of our
theory.

\section{Conclusion}
\label{sec:conclusion}

The modifications of the superconducting properties induced by strain in
Nb$_3$Sn, the most promising candidate for the next-generation high-power
magnets, has major practical consequences. To explain this strain dependence, we
have developed a theory which emphasizes the importance of anharmonicity in the
deformation potential, as was put forward by Marckievicz \cite{Markiewicz-2004}.
The theory is build upon the assumption that the strain-induced modifications of
the superconducting properties are mainly due to a widening of the
electron-phonon spectral function, which reduces the electron-phonon coupling
responsible for superconductivity. This widening is attributed to the nonlinear
generation of sum- and difference-frequency phonons by the anharmonic terms in
the elastic energy. We have expressed the strain function, which gives the
dependence of the upper critical field on strain, as a function of the
anharmonic energy. This function is exponential in the strain, as recent
investigations have suggested \cite{Bordini-2013}. The theory can fit
critical-field data measured on Nb$_3$Sn wires, and allowed us to extract the
anharmonic energy from these measurements. This anharmonic energy compares
favorably with first-principles calculations for strained Nb$_3$Sn.

The model indicates that the pre-strain state of the superconductor plays a big
role in lowering the maximum critical field. Any reduction of the pre-strain is
expected to produce an increase of $B_{c2}$. Furthermore, in the strands
considered, the superconductor seems to be in a pre-strain state involving a
tensile hydrostatic part of the order of 0.3--0.4\%, as well as a deviatoric
part. In the strands produced following the bronze route, the hydrostatic
pre-strain is nearly cancelled, in the direction of the strand, by a
longitudinal compressive term, while in the transverse directions the net strain
is a tensile one, of the order of 0.45\%. For strands produced by the
internal-tin and powder-in-tube methods, the model suggests that the net
pre-strain is less anisotropic, typically 0.1--0.3\% (respectively, 0.3--0.5\%)
in the longitudinal (respectively, transverse) directions. The more isotropic
pre-strain of the latter wires may explain their tendency to exhibit slightly
larger critical fields. More generally, our theory suggest that any treatment or
engineering step leading to a reduction of the anharmonicity, would result in an
improvement of the superconducting properties under strain.

\acknowledgments

We are grateful to C. Senatore, G. Mondonico, and T. Jarlborg for useful
discussions. This work was supported by the Swiss National Science Foundation
through Division II and MaNEP.

\appendix

\section{\boldmath Elements of nonlinear elasticity, and theoretical elastic
constants of bulk Nb$_3$Sn}
\label{app:elastic-constants}

The theory of nonlinear elasticity, and the propagation of waves in a nonlinear
elastic medium, have been described many times. In order to fix the notations,
we review here the elements of the theory which are relevant for our study. The
elastic energy density of an isotropic medium, up to third order in the strain
tensor $\varepsilon_{ij}$, can be expressed as \cite{Landau-1986, Murnaghan-1951}
	\begin{equation}\label{eq:elastic-energy-1}
		U=\frac{\lambda+2\mu}{2}\,I_1^2-2\mu\,I_2+\frac{\ell+2m}{3}\,I_1^3-2m\,I_1I_2+n\,I_3.
	\end{equation}
$\lambda$ and $\mu$ are the second-order elastic constants (Lam{\'e}
coefficients), $\ell$, $m$, and $n$ are the third-order Murnaghan coefficients,
and $I_{1,2,3}$ are three invariants of the strain tensor,
$I_1=\varepsilon_{ii}$ (the usual index summation convention is used),
$I_2=\varepsilon_{11}\varepsilon_{22}-\varepsilon_{12}\varepsilon_{21}+
\varepsilon_{22}\varepsilon_{33}-\varepsilon_{23}\varepsilon_{32}+
\varepsilon_{33}\varepsilon_{11}-\varepsilon_{31}\varepsilon_{13}$, and
$I_3=\det\varepsilon_{ij}$. For a deformation represented by the displacement
field $\vec{u}(\vec{x})$, the components of the strain tensor are
	\begin{equation}
		\varepsilon_{ij}=\frac{1}{2}\left(\frac{\partial u_i}{\partial x_j}
		+\frac{\partial u_j}{\partial x_i}+\frac{\partial u_k}{\partial x_i}
		\frac{\partial u_k}{\partial x_j}\right).
	\end{equation}
The elastic waves obey the equation of motion
$\rho\ddot{u}_i=\partial\sigma_{ij}/\partial x_j$, where $\rho$ is the mass
density, and $\sigma_{ij}$ is the stress tensor. Without attenuation and
external driving forces, the stress tensor is $\sigma_{ij}=\partial
U/\partial(\partial u_i/\partial x_j)$. For a longitudinal plane wave of
amplitude $A$ propagating in the direction $x$,
$\vec{u}(\vec{x},t)=\hat{\vec{x}}A(x,t)$, and ignoring terms of order three and
above in the displacement field, the equation of motion becomes
	\begin{equation}\label{eq:waveA}
		\frac{\partial^2A}{\partial x^2}-\frac{1}{v^2}\frac{\partial^2A}{\partial t^2}
		=-b\frac{\partial}{\partial x}\left(\frac{\partial A}{\partial x}\right)^2,
	\end{equation}
where the propagation velocity is given by $\rho v^2=\lambda+2\mu$, and the
nonlinear factor is
	\begin{equation}\label{eq:b}
		b=\frac{3}{2}+\frac{\ell+2m}{\lambda+2\mu}.
	\end{equation}
We discuss in Appendix~\ref{app:sum-and-difference} a particular solution of
Eq.~(\ref{eq:waveA}), which describes, starting from two initial waves, the
nonlinear generation of a third wave with sum frequency.

In order to estimate the elastic coefficients of Nb$_3$Sn, we have performed
first-principles electronic-structure calculations using {\sc Quantum ESPRESSO}
\cite{QE-2009}, and ultra-soft pseudo-potentials.\footnote{The pseudo-potential
files are Nb.pw91-nsp-van.UPF and Sn.pw91-n-van.UPF from
\href{http://www.quantum-espresso.org}{www.quantum-espresso.org}} We used a
Monkhorst-Pack mesh of $12\times12\times12$ $k$-points for the Brillouin-zone
integrations, with a 0.002~Rydberg gaussian broadening of the levels. The
wave-function cutoff was set to 40~Rydberg.

\begin{figure}[tb]
\includegraphics[width=\columnwidth]{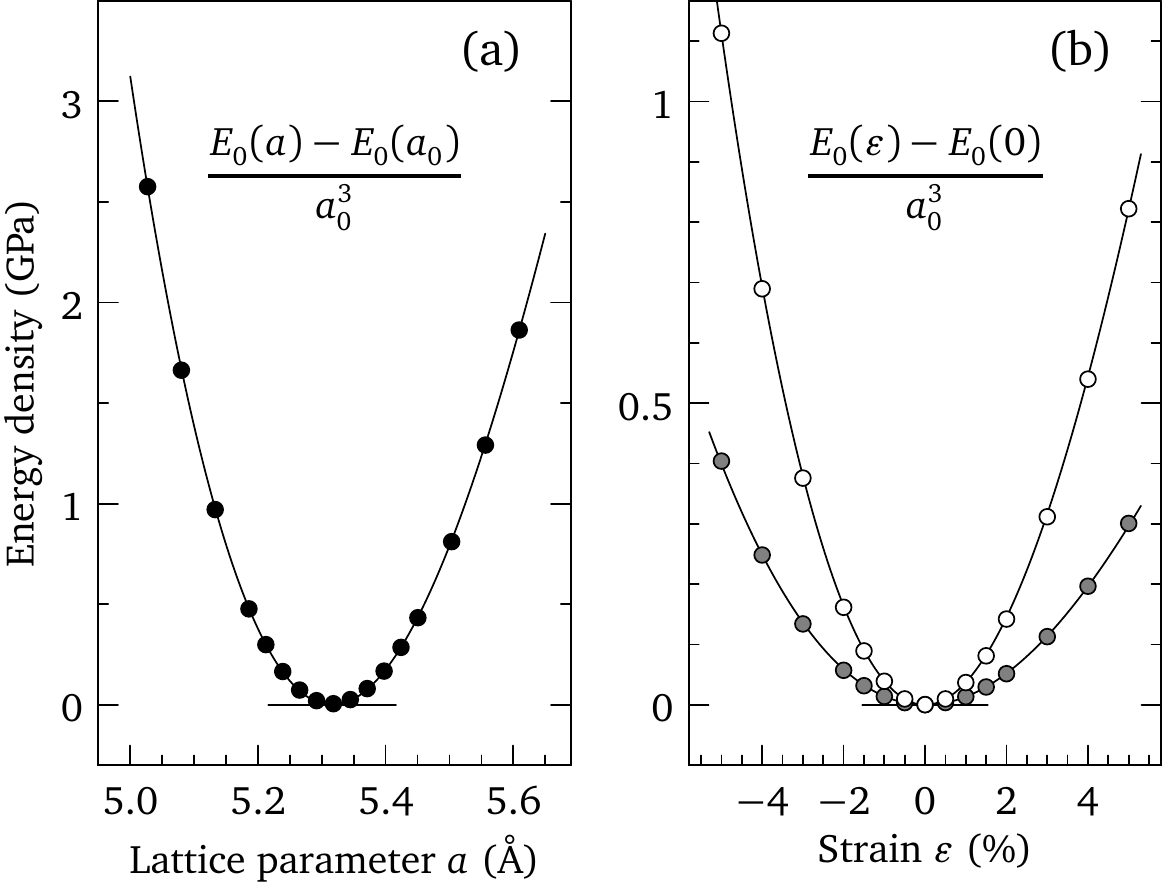}
\caption{\label{fig:fig3}
Elastic energy of Nb$_3$Sn for (a) hydrostatic strain and (b) longitudinal (gray
dots) and transverse (white dots) strain. The dots are the first-principles
results, the lines are fits to Eq.~(\ref{eq:elastic-energy-1}).
}
\end{figure}

The total energy of the cubic crystal is plotted in Fig.~\ref{fig:fig3}(a) as a
function of the lattice parameter. For hydrostatic strain, i.e.,
$\varepsilon_{ij}=\delta_{ij}\varepsilon$, the formula
(\ref{eq:elastic-energy-1}) reduces to the simple form
$U=(9K/2)\varepsilon^2+(9\ell+n)\varepsilon^3$, where $K=\lambda+2\mu/3$ is the
bulk modulus. This form, with $\varepsilon=a/a_0-1$, can fit the data in
Fig.~\ref{fig:fig3}(a) very well, and yields an equilibrium lattice parameter
$a_0=5.32$~\AA, a bulk modulus $K=165$~GPa, and the relation
$9\ell+n=-2368$~GPa. As Nb$_3$Sn is not isotropic, these coefficients should be
regarded as effective parameters used to represent the elastic energy of the
material. The theoretical lattice parameter is $0.8\%$ larger than the
low-temperature experimental value of $5.28$~\AA\ reported in
Ref.~\onlinecite{Muzzi-2012}. The theoretical bulk modulus compares very well
with the experimental value of 163~GPa reported in
Ref.~\onlinecite{Testardi-1973}. In order to determine the remaining
coefficients, we have calculated the energy for two non-hydrostatic diagonal
strains, either longitudinal ($\varepsilon_{11}=\varepsilon$,
$\varepsilon_{22}=\varepsilon_{33}=0$) or transverse ($\varepsilon_{11}=0$,
$\varepsilon_{22}=\varepsilon_{33}=\varepsilon$). The results of the calculation
and the fits are shown in Fig.~\ref{fig:fig3}(b). In the longitudinal case, the
elastic energy (\ref{eq:elastic-energy-1}) is
$U=(3K/2)(1-\nu)/(1+\nu)\varepsilon^2+(1/3)(\ell+2m)\varepsilon^3$, where
$\nu=(\lambda/2)/(\lambda+\mu)$ is the Poisson ratio, while in the transverse
case we have $U=3K/(1+\nu)\varepsilon^2+(4/3)(2\ell+m)\varepsilon^3$. Using the
value of $K$ obtained above, the fits yield $\nu=0.28$, $\ell+2m=-1232$~GPa, and
$2\ell+m=-874$~GPa. The resulting Murnaghan coefficients are therefore
$\ell=-172$, $m=-530$, and $n=-820$~GPa.

The theoretical coefficients correspond to a Young modulus
$E=3K(1-2\nu)=218$~GPa, significantly larger than the value $\sim130$~GPa
measured for Nb$_3$Sn wires. The theoretical Poisson ratio is also appreciably
lower than the experimental value of 0.36. These numbers can not be directly
compared, however, because the experiments are performed on wires in which the
Nb$_3$Sn material is strained, in addition to being mixed with other materials
forming the structure. In order to clarify the role of the pre-strain on the
elastic properties of Nb$_3$Sn, we have calculated the elastic energy in the
stain configuration considered in Sec.~\ref{sec:pre-strained-wires}, i.e.,
$\varepsilon_{11}=\varepsilon_0-\varepsilon_a^*+\varepsilon_a$,
$\varepsilon_{22}=\varepsilon_{33}=\varepsilon_0+\nu\varepsilon_a^*
-\nu\varepsilon_a$. We used the values $\varepsilon_0=0.34\%$,
$\varepsilon_a^*=0.33\%$ (see Table~\ref{tab:fits}) and the experimental Poisson
ratio $\nu=0.36$. The resulting elastic energy (Fig.~\ref{fig:fig2}) is not well
described by Eq.~(\ref{eq:elastic-energy-1}), if the equilibrium elastic
constants are used, but can be well fitted using effective parameters
corresponding to the experimental values (see Sec.~\ref{sec:fits}).

We close this appendix with the expression of the elastic energy
(\ref{eq:elastic-energy-1}) in terms of the invariants $I_1$, $J_2$, and $J_3$
introduced in Sec.~\ref{sec:pre-strained-wires}. These invariants are functions
of the infinitesimal strains $\varepsilon_i=\partial u_i/\partial x_i$ along the
principal axes, which are related to the diagonal strains $\varepsilon_{ii}$ by
$\varepsilon_{ii}=\varepsilon_i+\varepsilon_i^2/2$. Inserting this in
Eq.~(\ref{eq:elastic-energy-1}), a sixth-order development in the infinitesimal
strains results, which we express up to fourth order by means of the invariants
$I_1$, $J_2$, and $J_3$:
	\begin{multline}\label{eq:elastic-energy-2}
		U=\frac{3\lambda+2\mu}{6}\,I_1^2+2\mu\,J_2
		+\frac{9\lambda+6\mu+18\ell+2n}{54}\,I_1^3\\
		+\frac{3\lambda+6\mu+6m-n}{3}\,I_1J_2+(3\mu+n)\,J_3\\
		+\frac{3\lambda+2\mu+36\ell+4n}{216}\,I_1^4
		+\frac{\lambda+2\mu+6\ell+10m-n}{6}\,I_1^2J_2\\
		+\frac{2\mu+6m+n}{2}\,I_1J_3+\frac{\lambda+\mu+4m}{2}\,J_2^2.
	\end{multline}
With the relations $\lambda=E\nu/[(1+\nu)(1-2\nu)]$ and $\mu=(E/2)/(1+\nu)$, the
second-order terms of (\ref{eq:elastic-energy-2}) reduce to
Eq.~(\ref{eq:harmonic_energy}).

\section{Nonlinear generation of acoustic waves}
\label{app:sum-and-difference}

A longitudinal plane wave propagating without attenuation in an isotropic
elastic solid obeys, to lowest order in the nonlinearity, the equation of motion
(\ref{eq:waveA}). $A(x,t)$ is the displacement amplitude in the direction $x$,
which is also the direction of propagation, and $v$ is the propagation velocity.
The nonlinear term in the wave equation gives rise to sum- and
difference-frequency generation. We shall only consider the case of sum
frequency for simplicity. In this process, two plane waves of amplitudes
$A_1(t)$ and $A_2(t)$, and frequencies $\omega_1$ and $\omega_2$, generate a
third wave of amplitude $A_3(t)$ and frequency $\omega_3=\omega_1+\omega_2$. We
assume a linear dispersion relation, $\omega=vq$, which means that the
phase-matching condition is satisfied if the generated wave has wave number
$q_3=q_1+q_2$. We therefore consider a solution of Eq.~(\ref{eq:waveA}) in the
form
	\begin{multline}\label{eq:three_waves}
		A(x,t)=\frac{1}{2}\left[A_1(t)e^{iq_1(x-vt)}+A_2(t)e^{iq_2(x-vt)}\right.\\
		\left. +A_3(t)e^{i(q_1+q_2)(x-vt)}+\mathrm{c.c.}\right].
	\end{multline}
The amplitudes $A_i(t)$ are slow functions of $t$, and the initial condition is
$A_3(0)=0$. Inserting the solution (\ref{eq:three_waves}) into the wave equation
(\ref{eq:waveA}), and neglecting terms of order $d^2A_i(t)/dt^2$, we obtain the
evolution of the amplitude for each of the three frequencies:
	\begin{subequations}\label{eq:waveAi}
	\begin{align}
		\label{eq:A1dot}
		dA_1(t)/dt&=-(bv/2)q_2q_3\,A_2^*(t)A_3^{ }(t)\\
		\label{eq:A2dot}
		dA_2(t)/dt&=-(bv/2)q_1q_3\,A_1^*(t)A_3^{ }(t)\\
		\label{eq:A3dot}
		dA_3(t)/dt&=+(bv/2)q_1q_2\,A_1^{ }(t)A_2^{ }(t).
	\end{align}
	\end{subequations}
Taking the time derivative of (\ref{eq:A3dot}), and using (\ref{eq:A1dot}) and
(\ref{eq:A2dot}), the following second-order equation results for $A_3(t)$:
	\begin{multline}\label{eq:A3dotdot}
		d^2A_3(t)/dt^2=-(b/2)^2\omega_1\omega_2\\
		\times\left[q_1q_3|A_1(t)|^2+q_2q_3|A_2(t)|^2\right]A_3(t).
	\end{multline}
In order to eliminate the amplitudes $A_1(t)$ and $A_2(t)$ from this equation,
we note that the quantity
	\begin{equation}
		C=q_1q_3|A_1(t)|^2+q_2q_3|A_2(t)|^2+2q_3^2|A_3(t)|^2
	\end{equation}
is conserved during the evolution governed by (\ref{eq:waveAi}). This allows us
to rewrite Eq.~(\ref{eq:A3dotdot}) as a differential equation for $A_3(t)$ only:
	\begin{equation}\label{eq:A3dotdot1}
		\frac{d^2A_3(t)}{dt^2}=-(b/2)^2\omega_1\omega_2
		\left(C-2q_3^2|A_3(t)|^2\right)A_3(t).
	\end{equation}
This nonlinear equation admits a remarkably simple solution, which satisfies the
initial condition $A_3(0)=0$, namely
	\begin{equation}
		A_3(t)=\frac{1}{q_3}\sqrt{\frac{C}{2}}
		\tanh\left(t\sqrt{\frac{C}{2}(b/2)^2\omega_1\omega_2}\right).
	\end{equation}
The generated amplitude increases first linearly with time, and then saturates.
The saturation amplitude is reached when the energy available in the two primary
waves has been completely transferred to the secondary wave. We therefore
identify the energy carried by the generated wave, when it reaches saturation,
with the anharmonic energy, $\anharmonicU$, that was initially available for
sum-frequency generation. Since the energy density of an harmonic wave of
frequency $\omega$ and amplitude $A$ is $\rho(\omega A)^2$, we obtain an
interpretation for the conserved quantity $C$, in terms of $\tilde{U}$: 
	\begin{equation}
		\tilde{U}=\rho[\omega_3\,A_3(\infty)]^2=\rho v^2(C/2).
	\end{equation}
This allows us to write the amplitude of the generated wave in the form quoted
in the main text,
	\begin{equation}
		\omega_3\,A_3(t)=\sqrt{\tilde{U}/\rho}
		\tanh\left(tq_{12}\sqrt{\tilde{U}/\rho}\right),
	\end{equation}
where $q_{12}=|b/2|\sqrt{q_1q_2}$.


\begin{thebibliography}{41}%
\makeatletter
\providecommand \@ifxundefined [1]{%
 \@ifx{#1\undefined}
}%
\providecommand \@ifnum [1]{%
 \ifnum #1\expandafter \@firstoftwo
 \else \expandafter \@secondoftwo
 \fi
}%
\providecommand \@ifx [1]{%
 \ifx #1\expandafter \@firstoftwo
 \else \expandafter \@secondoftwo
 \fi
}%
\providecommand \natexlab [1]{#1}%
\providecommand \enquote  [1]{``#1''}%
\providecommand \bibnamefont  [1]{#1}%
\providecommand \bibfnamefont [1]{#1}%
\providecommand \citenamefont [1]{#1}%
\providecommand \href@noop [0]{\@secondoftwo}%
\providecommand \href [0]{\begingroup \@sanitize@url \@href}%
\providecommand \@href[1]{\@@startlink{#1}\@@href}%
\providecommand \@@href[1]{\endgroup#1\@@endlink}%
\providecommand \@sanitize@url [0]{\catcode `\\12\catcode `\$12\catcode
  `\&12\catcode `\#12\catcode `\^12\catcode `\_12\catcode `\%12\relax}%
\providecommand \@@startlink[1]{}%
\providecommand \@@endlink[0]{}%
\providecommand \url  [0]{\begingroup\@sanitize@url \@url }%
\providecommand \@url [1]{\endgroup\@href {#1}{\urlprefix }}%
\providecommand \urlprefix  [0]{URL }%
\providecommand \Eprint [0]{\href }%
\providecommand \doibase [0]{http://dx.doi.org/}%
\providecommand \selectlanguage [0]{\@gobble}%
\providecommand \bibinfo  [0]{\@secondoftwo}%
\providecommand \bibfield  [0]{\@secondoftwo}%
\providecommand \translation [1]{[#1]}%
\providecommand \BibitemOpen [0]{}%
\providecommand \bibitemStop [0]{}%
\providecommand \bibitemNoStop [0]{.\EOS\space}%
\providecommand \EOS [0]{\spacefactor3000\relax}%
\providecommand \BibitemShut  [1]{\csname bibitem#1\endcsname}%
\let\auto@bib@innerbib\@empty
\bibitem [{\citenamefont {Miyazaki}\ \emph {et~al.}(1999)\citenamefont
  {Miyazaki}, \citenamefont {Murakami}, \citenamefont {Hase}, \citenamefont
  {Shimada}, \citenamefont {Itoh}, \citenamefont {Kiyoshi}, \citenamefont
  {Takeuchi}, \citenamefont {Inoue},\ and\ \citenamefont
  {Wada}}]{Miyazaki-1999}%
  \BibitemOpen
  \bibfield  {author} {\bibinfo {author} {\bibfnamefont {T.}~\bibnamefont
  {Miyazaki}}, \bibinfo {author} {\bibfnamefont {Y.}~\bibnamefont {Murakami}},
  \bibinfo {author} {\bibfnamefont {T.}~\bibnamefont {Hase}}, \bibinfo {author}
  {\bibfnamefont {M.}~\bibnamefont {Shimada}}, \bibinfo {author} {\bibfnamefont
  {K.}~\bibnamefont {Itoh}}, \bibinfo {author} {\bibfnamefont {T.}~\bibnamefont
  {Kiyoshi}}, \bibinfo {author} {\bibfnamefont {T.}~\bibnamefont {Takeuchi}},
  \bibinfo {author} {\bibfnamefont {K.}~\bibnamefont {Inoue}}, \ and\ \bibinfo
  {author} {\bibfnamefont {H.}~\bibnamefont {Wada}},\ }\href 
  {\doibase 10.1109/77.784991} {\bibfield  {journal} {\bibinfo  {journal} {IEEE Trans.
  Appl. Supercond.}\ }\textbf {\bibinfo {volume} {9}},\ \bibinfo {pages} {2505}
  (\bibinfo {year} {1999})}\BibitemShut {NoStop}%
\bibitem [{\citenamefont {Mitchell}\ \emph {et~al.}(2009)\citenamefont
  {Mitchell}, \citenamefont {Bauer}, \citenamefont {Bessette}, \citenamefont
  {Devred}, \citenamefont {Gallix}, \citenamefont {Jong}, \citenamefont
  {Knaster}, \citenamefont {Libeyre}, \citenamefont {Lim}, \citenamefont
  {Sahu},\ and\ \citenamefont {Simon}}]{Mitchell-2009}%
  \BibitemOpen
  \bibfield  {author} {\bibinfo {author} {\bibfnamefont {N.}~\bibnamefont
  {Mitchell}}, \bibinfo {author} {\bibfnamefont {P.}~\bibnamefont {Bauer}},
  \bibinfo {author} {\bibfnamefont {D.}~\bibnamefont {Bessette}}, \bibinfo
  {author} {\bibfnamefont {A.}~\bibnamefont {Devred}}, \bibinfo {author}
  {\bibfnamefont {R.}~\bibnamefont {Gallix}}, \bibinfo {author} {\bibfnamefont
  {C.}~\bibnamefont {Jong}}, \bibinfo {author} {\bibfnamefont {J.}~\bibnamefont
  {Knaster}}, \bibinfo {author} {\bibfnamefont {P.}~\bibnamefont {Libeyre}},
  \bibinfo {author} {\bibfnamefont {B.}~\bibnamefont {Lim}}, \bibinfo {author}
  {\bibfnamefont {A.}~\bibnamefont {Sahu}}, \ and\ \bibinfo {author}
  {\bibfnamefont {F.}~\bibnamefont {Simon}},\ }\href 
  {\doibase 10.1016/j.fusengdes.2009.01.006} {\bibfield  {journal} {\bibinfo  {journal}
  {Fusion Engineering and Design}\ }\textbf {\bibinfo {volume} {84}},\ \bibinfo
  {pages} {113} (\bibinfo {year} {2009})}\BibitemShut {NoStop}%
\bibitem [{\citenamefont {Bottura}\ \emph {et~al.}(2012)\citenamefont
  {Bottura}, \citenamefont {de~Rijk}, \citenamefont {Rossi},\ and\
  \citenamefont {Todesco}}]{Bottura-2012}%
  \BibitemOpen
  \bibfield  {author} {\bibinfo {author} {\bibfnamefont {L.}~\bibnamefont
  {Bottura}}, \bibinfo {author} {\bibfnamefont {G.}~\bibnamefont {de~Rijk}},
  \bibinfo {author} {\bibfnamefont {L.}~\bibnamefont {Rossi}}, \ and\ \bibinfo
  {author} {\bibfnamefont {E.}~\bibnamefont {Todesco}},\ }\href 
  {\doibase 10.1109/TASC.2012.2186109} {\bibfield  {journal} {\bibinfo  {journal} {IEEE
  Trans. Appl. Supercond.}\ }\textbf {\bibinfo {volume} {22}},\ \bibinfo
  {pages} {4002008} (\bibinfo {year} {2012})}\BibitemShut {NoStop}%
\bibitem [{\citenamefont {Arbman}\ and\ \citenamefont
  {Jarlborg}(1978)}]{Arbman-1978}%
  \BibitemOpen
  \bibfield  {author} {\bibinfo {author} {\bibfnamefont {G.}~\bibnamefont
  {Arbman}}\ and\ \bibinfo {author} {\bibfnamefont {T.}~\bibnamefont
  {Jarlborg}},\ }\href {\doibase 10.1016/0038-1098(78)90759-7} {\bibfield
  {journal} {\bibinfo  {journal} {Sol. State Com.}\ }\textbf {\bibinfo {volume}
  {26}},\ \bibinfo {pages} {857} (\bibinfo {year} {1978})}\BibitemShut
  {NoStop}%
\bibitem [{\citenamefont {Klein}\ \emph {et~al.}(1979)\citenamefont {Klein},
  \citenamefont {Boyer},\ and\ \citenamefont
  {Papaconstantopoulos}}]{Klein-1979}%
  \BibitemOpen
  \bibfield  {author} {\bibinfo {author} {\bibfnamefont {B.~M.}\ \bibnamefont
  {Klein}}, \bibinfo {author} {\bibfnamefont {L.~L.}\ \bibnamefont {Boyer}}, \
  and\ \bibinfo {author} {\bibfnamefont {D.~A.}\ \bibnamefont
  {Papaconstantopoulos}},\ }\href {\doibase 10.1103/PhysRevLett.42.530}
  {\bibfield  {journal} {\bibinfo  {journal} {Phys. Rev. Lett.}\ }\textbf
  {\bibinfo {volume} {42}},\ \bibinfo {pages} {530} (\bibinfo {year}
  {1979})}\BibitemShut {NoStop}%
\bibitem [{\citenamefont {Paduani}(2009)}]{Paduani-2009}%
  \BibitemOpen
  \bibfield  {author} {\bibinfo {author} {\bibfnamefont {C.}~\bibnamefont
  {Paduani}},\ }\href {\doibase 10.1016/j.ssc.2009.05.011} {\bibfield
  {journal} {\bibinfo  {journal} {Sol. State Com.}\ }\textbf {\bibinfo {volume}
  {149}},\ \bibinfo {pages} {1269} (\bibinfo {year} {2009})}\BibitemShut
  {NoStop}%
\bibitem [{\citenamefont {Gor'kov}\ and\ \citenamefont
  {Dorokhov}(1976)}]{Gorkov-1976}%
  \BibitemOpen
  \bibfield  {author} {\bibinfo {author} {\bibfnamefont {L.~P.}\ \bibnamefont
  {Gor'kov}}\ and\ \bibinfo {author} {\bibfnamefont {O.~N.}\ \bibnamefont
  {Dorokhov}},\ }\href {\doibase 10.1007/BF00655212} {\bibfield  {journal}
  {\bibinfo  {journal} {J. Low Temp. Phys.}\ }\textbf {\bibinfo {volume}
  {22}},\ \bibinfo {pages} {1} (\bibinfo {year} {1976})}\BibitemShut {NoStop}%
\bibitem [{\citenamefont {Ekin}(1980)}]{Ekin-1980}%
  \BibitemOpen
  \bibfield  {author} {\bibinfo {author} {\bibfnamefont {J.~W.}\ \bibnamefont
  {Ekin}},\ }\href {\doibase 10.1016/0011-2275(80)90191-5} {\bibfield
  {journal} {\bibinfo  {journal} {Cryogenics}\ }\textbf {\bibinfo {volume}
  {20}},\ \bibinfo {pages} {611} (\bibinfo {year} {1980})}\BibitemShut
  {NoStop}%
\bibitem [{\citenamefont {Taylor}\ and\ \citenamefont
  {Hampshire}(2005)}]{Taylor-2005}%
  \BibitemOpen
  \bibfield  {author} {\bibinfo {author} {\bibfnamefont {D.~M.~J.}\
  \bibnamefont {Taylor}}\ and\ \bibinfo {author} {\bibfnamefont {D.~P.}\
  \bibnamefont {Hampshire}},\ }\href {\doibase 10.1088/0953-2048/18/12/005}
  {\bibfield  {journal} {\bibinfo  {journal} {Supercond. Sci. Technol.}\
  }\textbf {\bibinfo {volume} {18}},\ \bibinfo {pages} {S241} (\bibinfo {year}
  {2005})}\BibitemShut {NoStop}%
\bibitem [{\citenamefont {ten Haken}\ \emph {et~al.}(1999)\citenamefont {ten
  Haken}, \citenamefont {Godeke},\ and\ \citenamefont {ten
  Kate}}]{tenHaken-1999}%
  \BibitemOpen
  \bibfield  {author} {\bibinfo {author} {\bibfnamefont {B.}~\bibnamefont {ten
  Haken}}, \bibinfo {author} {\bibfnamefont {A.}~\bibnamefont {Godeke}}, \ and\
  \bibinfo {author} {\bibfnamefont {H.~H.~J.}\ \bibnamefont {ten Kate}},\
  }\href {\doibase 10.1063/1.369667} {\bibfield  {journal} {\bibinfo  {journal}
  {J. App. Phys.}\ }\textbf {\bibinfo {volume} {85}},\ \bibinfo {pages} {3247}
  (\bibinfo {year} {1999})}\BibitemShut {NoStop}%
\bibitem [{\citenamefont {Godeke}(2005)}]{Godeke-2005}%
  \BibitemOpen
  \bibfield  {author} {\bibinfo {author} {\bibfnamefont {A.}~\bibnamefont
  {Godeke}},\ }\emph {\bibinfo {title} {Performance boundaries in Nb$_3$Sn
  superconductors}},\ \href@noop {} {Ph.D. thesis},\ \bibinfo  {school}
  {University of Twente}, \bibinfo {address} {Enschede, The Netherlands}
  (\bibinfo {year} {2005})\BibitemShut {NoStop}%
\bibitem [{\citenamefont {Testardi}(1971)}]{Testardi-1971}%
  \BibitemOpen
  \bibfield  {author} {\bibinfo {author} {\bibfnamefont {L.~R.}\ \bibnamefont
  {Testardi}},\ }\href {\doibase 10.1103/PhysRevB.3.95} {\bibfield  {journal}
  {\bibinfo  {journal} {Phys. Rev. B}\ }\textbf {\bibinfo {volume} {3}},\
  \bibinfo {pages} {95} (\bibinfo {year} {1971})}\BibitemShut {NoStop}%
\bibitem [{\citenamefont {Markiewicz}(2004)}]{Markiewicz-2004}%
  \BibitemOpen
  \bibfield  {author} {\bibinfo {author} {\bibfnamefont {W.~D.}\ \bibnamefont
  {Markiewicz}},\ }\href {\doibase 10.1016/j.cryogenics.2004.03.019} {\bibfield
   {journal} {\bibinfo  {journal} {Cryogenics}\ }\textbf {\bibinfo {volume}
  {44}},\ \bibinfo {pages} {767} (\bibinfo {year} {2004})}\BibitemShut
  {NoStop}%
\bibitem [{\citenamefont {Bordini}\ \emph {et~al.}(2013)\citenamefont
  {Bordini}, \citenamefont {Alknes}, \citenamefont {Bottura}, \citenamefont
  {Rossi},\ and\ \citenamefont {Valentinis}}]{Bordini-2013}%
  \BibitemOpen
  \bibfield  {author} {\bibinfo {author} {\bibfnamefont {B.}~\bibnamefont
  {Bordini}}, \bibinfo {author} {\bibfnamefont {P.}~\bibnamefont {Alknes}},
  \bibinfo {author} {\bibfnamefont {L.}~\bibnamefont {Bottura}}, \bibinfo
  {author} {\bibfnamefont {L.}~\bibnamefont {Rossi}}, \ and\ \bibinfo {author}
  {\bibfnamefont {D.}~\bibnamefont {Valentinis}},\ }\href 
  {\doibase 10.1088/0953-2048/26/7/075014} {\bibfield  {journal} {\bibinfo  {journal}
  {Supercond. Sci. Technol.}\ }\textbf {\bibinfo {volume} {26}},\ \bibinfo
  {pages} {075014} (\bibinfo {year} {2013})}\BibitemShut {NoStop}%
\bibitem [{\citenamefont {Valentinis}(2012)}]{Valentinis-2012}%
  \BibitemOpen
  \bibfield  {author} {\bibinfo {author} {\bibfnamefont {D.~F.}\ \bibnamefont
  {Valentinis}},\ }\emph {\bibinfo {title} {Anharmonic model of the strain
  sensitivity of low temperature superconductors}},\ \href@noop {} {Master's
  thesis},\ \bibinfo  {school} {Polytechnic of Milan}, \bibinfo {address}
  {Milano} (\bibinfo {year} {2012})\BibitemShut {NoStop}%
\bibitem [{\citenamefont {Bordini}\ \emph {et~al.}(2012)\citenamefont
  {Bordini}, \citenamefont {Bottura}, \citenamefont {Mondonico}, \citenamefont
  {Oberli}, \citenamefont {Richter}, \citenamefont {Seeber}, \citenamefont
  {Senatore}, \citenamefont {Takala},\ and\ \citenamefont
  {Valentinis}}]{Bordini-2012}%
  \BibitemOpen
  \bibfield  {author} {\bibinfo {author} {\bibfnamefont {B.}~\bibnamefont
  {Bordini}}, \bibinfo {author} {\bibfnamefont {L.}~\bibnamefont {Bottura}},
  \bibinfo {author} {\bibfnamefont {G.}~\bibnamefont {Mondonico}}, \bibinfo
  {author} {\bibfnamefont {L.}~\bibnamefont {Oberli}}, \bibinfo {author}
  {\bibfnamefont {D.}~\bibnamefont {Richter}}, \bibinfo {author} {\bibfnamefont
  {B.}~\bibnamefont {Seeber}}, \bibinfo {author} {\bibfnamefont
  {C.}~\bibnamefont {Senatore}}, \bibinfo {author} {\bibfnamefont
  {E.}~\bibnamefont {Takala}}, \ and\ \bibinfo {author} {\bibfnamefont
  {D.}~\bibnamefont {Valentinis}},\ }\href {\doibase 10.1109/TASC.2011.2178217}
  {\bibfield  {journal} {\bibinfo  {journal} {IEEE Trans. Appl. Supercond.}\
  }\textbf {\bibinfo {volume} {22}},\ \bibinfo {pages} {6000304} (\bibinfo
  {year} {2012})}\BibitemShut {NoStop}%
\bibitem [{\citenamefont {Scalapino}\ \emph {et~al.}(1966)\citenamefont
  {Scalapino}, \citenamefont {Schrieffer},\ and\ \citenamefont
  {Wilkins}}]{Scalapino-1966}%
  \BibitemOpen
  \bibfield  {author} {\bibinfo {author} {\bibfnamefont {D.~J.}\ \bibnamefont
  {Scalapino}}, \bibinfo {author} {\bibfnamefont {J.~R.}\ \bibnamefont
  {Schrieffer}}, \ and\ \bibinfo {author} {\bibfnamefont {J.~W.}\ \bibnamefont
  {Wilkins}},\ }\href {\doibase 10.1103/PhysRev.148.263} {\bibfield  {journal}
  {\bibinfo  {journal} {Phys. Rev.}\ }\textbf {\bibinfo {volume} {148}},\
  \bibinfo {pages} {263} (\bibinfo {year} {1966})}\BibitemShut {NoStop}%
\bibitem [{\citenamefont {McMillan}(1968)}]{McMillan-1968}%
  \BibitemOpen
  \bibfield  {author} {\bibinfo {author} {\bibfnamefont {W.~L.}\ \bibnamefont
  {McMillan}},\ }\href {\doibase 10.1103/PhysRev.167.331} {\bibfield  {journal}
  {\bibinfo  {journal} {Phys. Rev.}\ }\textbf {\bibinfo {volume} {167}},\
  \bibinfo {pages} {331} (\bibinfo {year} {1968})}\BibitemShut {NoStop}%
\bibitem [{\citenamefont {Dynes}(1972)}]{Dynes-1972}%
  \BibitemOpen
  \bibfield  {author} {\bibinfo {author} {\bibfnamefont {R.~C.}\ \bibnamefont
  {Dynes}},\ }\href {\doibase 10.1016/0038-1098(72)90603-5} {\bibfield
  {journal} {\bibinfo  {journal} {Sol. State Com.}\ }\textbf {\bibinfo {volume}
  {10}},\ \bibinfo {pages} {615} (\bibinfo {year} {1972})}\BibitemShut
  {NoStop}%
\bibitem [{\citenamefont {Allen}\ and\ \citenamefont
  {Dynes}(1975)}]{Allen-1975}%
  \BibitemOpen
  \bibfield  {author} {\bibinfo {author} {\bibfnamefont {P.~B.}\ \bibnamefont
  {Allen}}\ and\ \bibinfo {author} {\bibfnamefont {R.~C.}\ \bibnamefont
  {Dynes}},\ }\href {\doibase 10.1103/PhysRevB.12.905} {\bibfield  {journal}
  {\bibinfo  {journal} {Phys. Rev. B}\ }\textbf {\bibinfo {volume} {12}},\
  \bibinfo {pages} {905} (\bibinfo {year} {1975})}\BibitemShut {NoStop}%
\bibitem [{\citenamefont {Morel}\ and\ \citenamefont
  {Anderson}(1962)}]{Morel-1962}%
  \BibitemOpen
  \bibfield  {author} {\bibinfo {author} {\bibfnamefont {P.}~\bibnamefont
  {Morel}}\ and\ \bibinfo {author} {\bibfnamefont {P.~W.}\ \bibnamefont
  {Anderson}},\ }\href {\doibase 10.1103/PhysRev.125.1263} {\bibfield
  {journal} {\bibinfo  {journal} {Phys. Rev.}\ }\textbf {\bibinfo {volume}
  {125}},\ \bibinfo {pages} {1263} (\bibinfo {year} {1962})}\BibitemShut
  {NoStop}%
\bibitem [{\citenamefont {De~Marzi}\ \emph {et~al.}(2013)\citenamefont
  {De~Marzi}, \citenamefont {Morici}, \citenamefont {Muzzi}, \citenamefont
  {della Corte},\ and\ \citenamefont {Buongiorno~Nardelli}}]{DeMarzi-2012}%
  \BibitemOpen
  \bibfield  {author} {\bibinfo {author} {\bibfnamefont {G.}~\bibnamefont
  {De~Marzi}}, \bibinfo {author} {\bibfnamefont {L.}~\bibnamefont {Morici}},
  \bibinfo {author} {\bibfnamefont {L.}~\bibnamefont {Muzzi}}, \bibinfo
  {author} {\bibfnamefont {A.}~\bibnamefont {della Corte}}, \ and\ \bibinfo
  {author} {\bibfnamefont {M.}~\bibnamefont {Buongiorno~Nardelli}},\ }\href
  {\doibase 10.1088/0953-8984/25/13/135702} {\bibfield  {journal} {\bibinfo
  {journal} {J. Phys.: Cond. Mat.}\ }\textbf {\bibinfo {volume} {25}},\
  \bibinfo {pages} {135702} (\bibinfo {year} {2013})}\BibitemShut {NoStop}%
\bibitem [{\citenamefont {Oh}\ and\ \citenamefont {Kim}(2006)}]{Oh-2006}%
  \BibitemOpen
  \bibfield  {author} {\bibinfo {author} {\bibfnamefont {S.}~\bibnamefont
  {Oh}}\ and\ \bibinfo {author} {\bibfnamefont {K.}~\bibnamefont {Kim}},\
  }\href {\doibase 10.1063/1.2170415} {\bibfield  {journal} {\bibinfo
  {journal} {J. App. Phys.}\ }\textbf {\bibinfo {volume} {99}},\ \bibinfo
  {pages} {033909} (\bibinfo {year} {2006})}\BibitemShut {NoStop}%
\bibitem [{\citenamefont {Godeke}(2006)}]{Godeke-2006}%
  \BibitemOpen
  \bibfield  {author} {\bibinfo {author} {\bibfnamefont {A.}~\bibnamefont
  {Godeke}},\ }\href {\doibase 10.1088/0953-2048/19/8/R02} {\bibfield
  {journal} {\bibinfo  {journal} {Supercond. Sci. Technol.}\ }\textbf {\bibinfo
  {volume} {19}},\ \bibinfo {pages} {R68} (\bibinfo {year} {2006})}\BibitemShut
  {NoStop}%
\bibitem [{\citenamefont {Wolf}\ \emph {et~al.}(1980)\citenamefont {Wolf},
  \citenamefont {Zasadzinski}, \citenamefont {Arnold}, \citenamefont {Moore},
  \citenamefont {Rowell},\ and\ \citenamefont {Beasley}}]{Wolf-1980}%
  \BibitemOpen
  \bibfield  {author} {\bibinfo {author} {\bibfnamefont {E.~L.}\ \bibnamefont
  {Wolf}}, \bibinfo {author} {\bibfnamefont {J.}~\bibnamefont {Zasadzinski}},
  \bibinfo {author} {\bibfnamefont {G.~B.}\ \bibnamefont {Arnold}}, \bibinfo
  {author} {\bibfnamefont {D.~F.}\ \bibnamefont {Moore}}, \bibinfo {author}
  {\bibfnamefont {J.~M.}\ \bibnamefont {Rowell}}, \ and\ \bibinfo {author}
  {\bibfnamefont {M.~R.}\ \bibnamefont {Beasley}},\ }\href 
  {\doibase 10.1103/PhysRevB.22.1214} {\bibfield  {journal} {\bibinfo  {journal} {Phys.
  Rev. B}\ }\textbf {\bibinfo {volume} {22}},\ \bibinfo {pages} {1214}
  (\bibinfo {year} {1980})}\BibitemShut {NoStop}%
\bibitem [{\citenamefont {Kwo}\ and\ \citenamefont {Geballe}(1981)}]{Kwo-1981}%
  \BibitemOpen
  \bibfield  {author} {\bibinfo {author} {\bibfnamefont {J.}~\bibnamefont
  {Kwo}}\ and\ \bibinfo {author} {\bibfnamefont {T.~H.}\ \bibnamefont
  {Geballe}},\ }\href {\doibase 10.1103/PhysRevB.23.3230} {\bibfield  {journal}
  {\bibinfo  {journal} {Phys. Rev. B}\ }\textbf {\bibinfo {volume} {23}},\
  \bibinfo {pages} {3230} (\bibinfo {year} {1981})}\BibitemShut {NoStop}%
\bibitem [{\citenamefont {Kihlstrom}\ \emph {et~al.}(1984)\citenamefont
  {Kihlstrom}, \citenamefont {Mael},\ and\ \citenamefont
  {Geballe}}]{Kihlstrom-1984}%
  \BibitemOpen
  \bibfield  {author} {\bibinfo {author} {\bibfnamefont {K.~E.}\ \bibnamefont
  {Kihlstrom}}, \bibinfo {author} {\bibfnamefont {D.}~\bibnamefont {Mael}}, \
  and\ \bibinfo {author} {\bibfnamefont {T.~H.}\ \bibnamefont {Geballe}},\
  }\href {\doibase 10.1103/PhysRevB.29.150} {\bibfield  {journal} {\bibinfo
  {journal} {Phys. Rev. B}\ }\textbf {\bibinfo {volume} {29}},\ \bibinfo
  {pages} {150} (\bibinfo {year} {1984})}\BibitemShut {NoStop}%
\bibitem [{\citenamefont {Delaire}\ \emph {et~al.}(2008)\citenamefont
  {Delaire}, \citenamefont {Lucas}, \citenamefont {Mu{\~n}oz}, \citenamefont
  {Kresch},\ and\ \citenamefont {Fultz}}]{Delaire-2008}%
  \BibitemOpen
  \bibfield  {author} {\bibinfo {author} {\bibfnamefont {O.}~\bibnamefont
  {Delaire}}, \bibinfo {author} {\bibfnamefont {M.~S.}\ \bibnamefont {Lucas}},
  \bibinfo {author} {\bibfnamefont {J.~A.}\ \bibnamefont {Mu{\~n}oz}}, \bibinfo
  {author} {\bibfnamefont {M.}~\bibnamefont {Kresch}}, \ and\ \bibinfo {author}
  {\bibfnamefont {B.}~\bibnamefont {Fultz}},\ }\href 
  {\doibase 10.1103/PhysRevLett.101.105504} {\bibfield  {journal} {\bibinfo  {journal}
  {Phys. Rev. Lett.}\ }\textbf {\bibinfo {volume} {101}},\ \bibinfo {pages}
  {105504} (\bibinfo {year} {2008})}\BibitemShut {NoStop}%
\bibitem [{\citenamefont {Boyd}(2003)}]{Boyd-2003}%
  \BibitemOpen
  \bibfield  {author} {\bibinfo {author} {\bibfnamefont {R.~W.}\ \bibnamefont
  {Boyd}},\ }\href@noop {} {\emph {\bibinfo {title} {Nonlinear Optics}}},\
  \bibinfo {edition} {3rd}\ ed.\ (\bibinfo  {publisher} {Academic Press},\
  \bibinfo {address} {Waltham},\ \bibinfo {year} {2003})\BibitemShut {NoStop}%
\bibitem [{\citenamefont {de~Gennes}(1999)}]{deGennes-1999}%
  \BibitemOpen
  \bibfield  {author} {\bibinfo {author} {\bibfnamefont {P.~G.}\ \bibnamefont
  {de~Gennes}},\ }\href@noop {} {\emph {\bibinfo {title} {Superconductivity of
  metals and alloys}}},\ Advanced book classics\ (\bibinfo  {publisher}
  {Perseus},\ \bibinfo {address} {Cambridge},\ \bibinfo {year}
  {1999})\BibitemShut {NoStop}%
\bibitem [{\citenamefont {ten Haken}(1994)}]{tenHaken-1994}%
  \BibitemOpen
  \bibfield  {author} {\bibinfo {author} {\bibfnamefont {B.}~\bibnamefont {ten
  Haken}},\ }\emph {\bibinfo {title} {Strain effects on the critical properties
  of high-field superconductors}},\ \href@noop {} {Ph.D. thesis},\ \bibinfo
  {school} {University of Twente}, \bibinfo {address} {Enschede, The
  Netherlands} (\bibinfo {year} {1994})\BibitemShut {NoStop}%
\bibitem [{\citenamefont {Muzzi}\ \emph {et~al.}(2012)\citenamefont {Muzzi},
  \citenamefont {Corato}, \citenamefont {della Corte}, \citenamefont
  {De~Marzi}, \citenamefont {Spina}, \citenamefont {Daniels}, \citenamefont
  {Di~Michiel}, \citenamefont {Buta}, \citenamefont {Mondonico}, \citenamefont
  {Seeber}, \citenamefont {Fl{\"u}kiger},\ and\ \citenamefont
  {Senatore}}]{Muzzi-2012}%
  \BibitemOpen
  \bibfield  {author} {\bibinfo {author} {\bibfnamefont {L.}~\bibnamefont
  {Muzzi}}, \bibinfo {author} {\bibfnamefont {V.}~\bibnamefont {Corato}},
  \bibinfo {author} {\bibfnamefont {A.}~\bibnamefont {della Corte}}, \bibinfo
  {author} {\bibfnamefont {G.}~\bibnamefont {De~Marzi}}, \bibinfo {author}
  {\bibfnamefont {T.}~\bibnamefont {Spina}}, \bibinfo {author} {\bibfnamefont
  {J.}~\bibnamefont {Daniels}}, \bibinfo {author} {\bibfnamefont
  {M.}~\bibnamefont {Di~Michiel}}, \bibinfo {author} {\bibfnamefont
  {F.}~\bibnamefont {Buta}}, \bibinfo {author} {\bibfnamefont {G.}~\bibnamefont
  {Mondonico}}, \bibinfo {author} {\bibfnamefont {B.}~\bibnamefont {Seeber}},
  \bibinfo {author} {\bibfnamefont {R.}~\bibnamefont {Fl{\"u}kiger}}, \ and\
  \bibinfo {author} {\bibfnamefont {C.}~\bibnamefont {Senatore}},\ }\href
  {\doibase 10.1088/0953-2048/25/5/054006} {\bibfield  {journal} {\bibinfo
  {journal} {Supercond. Sci. Technol.}\ }\textbf {\bibinfo {volume} {25}},\
  \bibinfo {pages} {054006} (\bibinfo {year} {2012})}\BibitemShut {NoStop}%
\bibitem [{\citenamefont {Freericks}\ \emph {et~al.}(2002)\citenamefont
  {Freericks}, \citenamefont {Liu}, \citenamefont {Quandt},\ and\ \citenamefont
  {Geerk}}]{Freericks-2002}%
  \BibitemOpen
  \bibfield  {author} {\bibinfo {author} {\bibfnamefont {J.~K.}\ \bibnamefont
  {Freericks}}, \bibinfo {author} {\bibfnamefont {A.~Y.}\ \bibnamefont {Liu}},
  \bibinfo {author} {\bibfnamefont {A.}~\bibnamefont {Quandt}}, \ and\ \bibinfo
  {author} {\bibfnamefont {J.}~\bibnamefont {Geerk}},\ }\href 
  {\doibase 10.1103/PhysRevB.65.224510} {\bibfield  {journal} {\bibinfo  {journal} {Phys.
  Rev. B}\ }\textbf {\bibinfo {volume} {65}},\ \bibinfo {pages} {224510}
  (\bibinfo {year} {2002})}\BibitemShut {NoStop}%
\bibitem [{\citenamefont {Mondonico}\ \emph {et~al.}(2010)\citenamefont
  {Mondonico}, \citenamefont {Seeber}, \citenamefont {Senatore}, \citenamefont
  {Fl{\"{u}}kiger}, \citenamefont {Corato}, \citenamefont {Marzi},\ and\
  \citenamefont {Muzzi}}]{Mondonico-2010}%
  \BibitemOpen
  \bibfield  {author} {\bibinfo {author} {\bibfnamefont {G.}~\bibnamefont
  {Mondonico}}, \bibinfo {author} {\bibfnamefont {B.}~\bibnamefont {Seeber}},
  \bibinfo {author} {\bibfnamefont {C.}~\bibnamefont {Senatore}}, \bibinfo
  {author} {\bibfnamefont {R.}~\bibnamefont {Fl{\"{u}}kiger}}, \bibinfo
  {author} {\bibfnamefont {V.}~\bibnamefont {Corato}}, \bibinfo {author}
  {\bibfnamefont {G.~D.}\ \bibnamefont {Marzi}}, \ and\ \bibinfo {author}
  {\bibfnamefont {L.}~\bibnamefont {Muzzi}},\ }\href 
  {\doibase 10.1063/1.3499649} {\bibfield  {journal} {\bibinfo  {journal} {J. App.
  Phys.}\ }\textbf {\bibinfo {volume} {108}},\ \bibinfo {pages} {093906}
  (\bibinfo {year} {2010})}\BibitemShut {NoStop}%
\bibitem [{\citenamefont {Guritanu}\ \emph {et~al.}(2004)\citenamefont
  {Guritanu}, \citenamefont {Goldacker}, \citenamefont {Bouquet}, \citenamefont
  {Wang}, \citenamefont {Lortz}, \citenamefont {Goll},\ and\ \citenamefont
  {Junod}}]{Guritanu-2004}%
  \BibitemOpen
  \bibfield  {author} {\bibinfo {author} {\bibfnamefont {V.}~\bibnamefont
  {Guritanu}}, \bibinfo {author} {\bibfnamefont {W.}~\bibnamefont {Goldacker}},
  \bibinfo {author} {\bibfnamefont {F.}~\bibnamefont {Bouquet}}, \bibinfo
  {author} {\bibfnamefont {Y.}~\bibnamefont {Wang}}, \bibinfo {author}
  {\bibfnamefont {R.}~\bibnamefont {Lortz}}, \bibinfo {author} {\bibfnamefont
  {G.}~\bibnamefont {Goll}}, \ and\ \bibinfo {author} {\bibfnamefont
  {A.}~\bibnamefont {Junod}},\ }\href {\doibase 10.1103/PhysRevB.70.184526}
  {\bibfield  {journal} {\bibinfo  {journal} {Phys. Rev. B}\ }\textbf {\bibinfo
  {volume} {70}},\ \bibinfo {pages} {184526} (\bibinfo {year}
  {2004})}\BibitemShut {NoStop}%
\bibitem [{\citenamefont {Axe}\ and\ \citenamefont {Shirane}(1973)}]{Axe-1973}%
  \BibitemOpen
  \bibfield  {author} {\bibinfo {author} {\bibfnamefont {J.~D.}\ \bibnamefont
  {Axe}}\ and\ \bibinfo {author} {\bibfnamefont {G.}~\bibnamefont {Shirane}},\
  }\href {\doibase 10.1103/PhysRevB.8.1965} {\bibfield  {journal} {\bibinfo
  {journal} {Phys. Rev. B}\ }\textbf {\bibinfo {volume} {8}},\ \bibinfo {pages}
  {1965} (\bibinfo {year} {1973})}\BibitemShut {NoStop}%
\bibitem [{\citenamefont {Landau}\ and\ \citenamefont
  {Lifshitz}(1986)}]{Landau-1986}%
  \BibitemOpen
  \bibfield  {author} {\bibinfo {author} {\bibfnamefont {L.~D.}\ \bibnamefont
  {Landau}}\ and\ \bibinfo {author} {\bibfnamefont {E.~M.}\ \bibnamefont
  {Lifshitz}},\ }\href@noop {} {\emph {\bibinfo {title} {Theory of
  Elasticity}}},\ \bibinfo {edition} {3rd}\ ed.\ (\bibinfo  {publisher}
  {Elsevier},\ \bibinfo {address} {Oxford},\ \bibinfo {year}
  {1986})\BibitemShut {NoStop}%
\bibitem [{\citenamefont {Murnaghan}(1951)}]{Murnaghan-1951}%
  \BibitemOpen
  \bibfield  {author} {\bibinfo {author} {\bibfnamefont {F.~D.}\ \bibnamefont
  {Murnaghan}},\ }\href@noop {} {\emph {\bibinfo {title} {Finite Deformation of
  an Elastic Solid}}}\ (\bibinfo  {publisher} {John Wiley \& Sons},\ \bibinfo
  {address} {New York},\ \bibinfo {year} {1951})\BibitemShut {NoStop}%
\bibitem [{\citenamefont {Giannozzi~\textit{et al.}}(2009)}]{QE-2009}%
  \BibitemOpen
  \bibfield  {author} {\bibinfo {author} {\bibfnamefont {P.}~\bibnamefont
  {Giannozzi~\textit{et al.}}},\ }\href 
  {\doibase 10.1088/0953-8984/21/39/395502} {\bibfield  {journal} {\bibinfo  {journal}
  {J. Phys.: Cond. Mat.}\ }\textbf {\bibinfo {volume} {21}},\ \bibinfo {pages}
  {395502} (\bibinfo {year} {2009})}\BibitemShut {NoStop}%
\bibitem [{Note1()}]{Note1}%
  \BibitemOpen
  \bibinfo {note} {The pseudo-potential files are Nb.pw91-nsp-van.UPF and
  Sn.pw91-n-van.UPF from \protect \href
  {http://www.quantum-espresso.org}{www.quantum-espresso.org}}\BibitemShut
  {NoStop}%
\bibitem [{\citenamefont {Testardi}(1973)}]{Testardi-1973}%
  \BibitemOpen
  \bibfield  {author} {\bibinfo {author} {\bibfnamefont {L.~R.}\ \bibnamefont
  {Testardi}},\ }\href {\doibase 10.1016/B978-0-12-477910-5.50009-X} {\bibfield
   {journal} {\bibinfo  {journal} {Phys. Acoust.}\ }\textbf {\bibinfo {volume}
  {10}},\ \bibinfo {pages} {193} (\bibinfo {year} {1973})}\BibitemShut
  {NoStop}%
\end{thebibliography}
\end{document}